\documentclass[preprint,12pt]{elsarticle}

\usepackage{graphicx}
\usepackage{amssymb}
\usepackage{amsmath}

\newcommand{\ld}{\lambda \Delta}

\newcommand{\rmd}{\textnormal{d}}
\newcommand{\rme}{\textnormal{e}}
\newtheorem{theorem}{Theorem}
\newtheorem{definition}{Definition}
\newtheorem{lemma}{Lemma}

\journal{BioSystems}

\begin{document}

\begin{frontmatter}

\title{Firing statistics of inhibitory neuron with delayed feedback.
II. Non-Markovian behavior.}

\author{K. G. Kravchuk}
\ead{kgkravchuk@bitp.kiev.ua}
\author{A. K. Vidybida}
\address{Bogolyubov Institute for Theoretical Physics, Metrologichna str. 14-B, 03680 Kyiv, Ukraine}
\ead{vidybida@bitp.kiev.ua}
\ead[url]{http://www.bitp.kiev.ua/pers/vidybida}


\begin{abstract}
The instantaneous state of a neural network consists of both the
degree of excitation of each neuron the network is composed of and positions of impulses in communication lines
between the neurons.
In neurophysiological experiments, the neuronal firing moments are registered, but not the state of communication lines.
But  future spiking moments depend essentially on the past positions of
impulses in the lines.
This suggests, that the sequence of intervals between firing moments (inter-spike intervals, ISIs)
in the network could be non-Markovian.

In this paper, we address this question for a simplest possible neural ``net'', namely,
a single inhibitory neuron with delayed feedback. The neuron
 receives excitatory input from the driving Poisson stream and inhibitory impulses from its own
output through the feedback line.
We obtain analytic expressions for conditional probability density $P(t_{n+1}\mid t_{n},\ldots,t_1,t_{0})$, 
which gives the probability to get an output
ISI of duration $t_{n+1}$ provided the previous $(n+1)$
 output ISIs had durations $t_{n},\ldots,t_1,t_{0}$.
It is proven exactly, that  $P(t_{n+1}\mid t_{n},\ldots,t_1,t_{0})$ does not
reduce to $P(t_{n+1}\mid t_{n},\ldots,t_{1})$ for any $n\ge0$. 
This means that the output ISIs stream cannot be represented as a Markov chain of any finite order.\smallskip
\end{abstract}

\end{frontmatter}

\section{Introduction}
\label{intro}

In a neural network, the main component parts are neurons and inter-neuronal 
communication lines -- axons  
\cite{Nicholls}.
 These same units are the main ones in most types of artificial neural networks \cite{Adeli}.
If so, then the instantaneous dynamical state of a network must include dynamical states of all the neurons and
communication lines the network is composed of. 
The state of a neuron can be described as its degree of 
excitation.
The state of a line consists of information of
whether the line is empty or conducts an impulse. 
If it does conduct, then the state of the line can be described by the amount of time
which is required for the impulse to reach the end of the line (time to live). 

In neurophysiological experiments, the triggering (spiking, firing) moments of individual neurons but not the states of communication lines  are registered.
The sequence of intervals between the consecutive moments (inter-spike intervals, ISIs) is frequently
considered as a renewal \cite{Holden} or Markovian \cite{Britvina} stochastic process. For a renewal process, 
the consecutive ISIs are mutually statistically independent. Moreover,
all statistical characteristics of a spike train must be derivable 
from the single-ISI probability distribution. 
Additionally, those characteristics must be the same for a shuffled
spike train, obtained by randomly reordering the ISIs, 
since shuffling does not change the single-ISI probability distribution.
On the other hand, the experimentally obtained spike trains in auditory \cite{Lowen}
and visual~\cite{Levine} sensory systems does not support the ISIs' mutual
independence. This is revealed by calculating the correlation coefficient between 
the adjacent ISIs, which appeared to be nonzero for the experimental spike trains, 
while it must be zero for any renewal process. 
Also, such characteristics as Fano factor
curve and firing rate distribution calculated for shuffled spike trains differ 
qualitatively from those obtained for the intact ones. These observations can be
associated with memory effects in the ISI sequence which arise from an underlying
non-renewal process. Recently \cite{RatnamNelson}, such a possibility 
was analyzed for weakly electric fish electrosensory afferents using 
high-order interval analysis, count analysis, and Markov-order analysis.
The authors conclude that the experimental evidence cannot reject 
the null hypothesis that the underlying Markov chain model 
is of order $m$ or higher, or maybe non-Markovian. 
The limited data sets used in \cite{RatnamNelson} allow to establish
a lower bound for $m$ as $m\ge7$ for some fibers.

What could be possible sources of such non-renewal, or even non-Markovian, 
behavior of ISI sequences in real neural network?
First, this behavior could be inherited from non-renewal (non-Markovian) character 
of the input signal.
Second, intrinsic neuronal properties, such as adaptation, could be responsible.
Finally, as we show here,
the presence of delayed feedback interconnections itself could
be the possible source of the non-Markovian behavior of ISI sequences.

The non-Markovian behavior of the ISI sequence 
from neuron in a network with delayed interconnections
is not surprising.
Indeed, 
the information about which neurons are spiking/silent
at any given moment of time leaves
unknown the position of impulses in the interconnection lines at that moment.
And it is the previous firing moments which determine the states 
of interconnection lines,
which in turn determine the next firing moments. 
Therefore, information about the previous neuronal 
firing moments
could improve our predicting ability as regards the next firing moments. 

In this paper, we consider a simplest neural ``net'', namely, a single 
inhibitory neuron with delayed feedback, which is driven with 
excitatory impulses from a Poisson process. 
As neuronal model we take binding neuron
as it allows rigorous mathematical treatment.
We study the ISI output stream of this system and prove that it
cannot be presented as Markovian chain of any finite order. 
This suggests that activity of a network, 
if presented in terms of neuronal interspike intervals, 
could be non-Markovian as well, provided the network includes components
with delayed interconnections, similar to that in the
Fig. \ref{fig:BNDF}.


\section{The object under consideration}
\subsection{Binding neuron model}
\label{sec:BN}
The understanding of mechanisms of higher brain functions expects a continuous reduction 
from higher activities to lower ones, eventually, to activities in individual neurons, 
expressed in terms of membrane potentials and ionic currents. 
But the description of the higher brain functions in terms of potentials and currents
in parts of individual neurons would be difficult, similarly as it would be difficult
to describe execution of computer programs  by a CPU in terms of Kirhgoff's laws.
In this connection, it would be helpful to abstract from the rules 
by which a neuron changes its membrane potentials to rules by which the input impulse signals 
are processed in the neuron and determine its output firing activity. The “coincidence detector”, and “temporal integrator” 
are the examples of such an abstraction, 
see discussion in \cite{Konig}.

\begin{figure}
\includegraphics[width=0.95\textwidth]{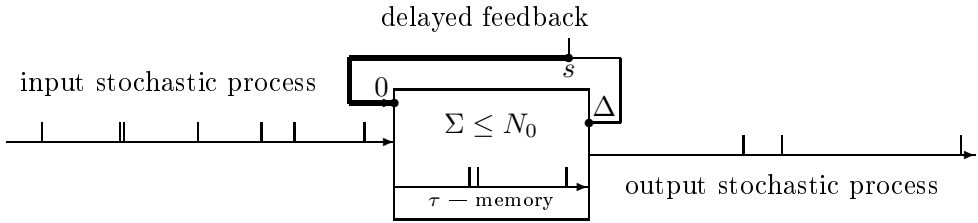}
\caption{Binding neuron with feedback line under Poisson stimulation. 
Multiple input lines with Poisson streams are joined into a single one here.
$\Delta$ is the delay duration in the feedback line, 
$s$ is the time left for the feedbacked impulse to reach the neuron.
}
\label{fig:BNDF}
\end{figure}

One more abstraction, the binding neuron (BN) model, 
is proposed as signal processing unit \cite{Vid96}, 
which can operate either as coincidence detector, or temporal integrator, 
depending on quantitative characteristics of stimulation applied. 
This conforms with behavior of real neurons, see, e.g. \cite{Rudolph,Lundstrom}.
The BN model 
describes functioning of a neuron in terms of discrete events, 
which are input and output impulses, and degree of temporal coherence between 
the input events, see \cite{Vid98} for detailed description.
Mathematically, this model can be realized as follows. We expect that all input 
impulses in all input lines are identical.
Each input impulse is stored in the BN for a fixed time, $\tau$. 
The $\tau$ is similar to the “tolerance interval” discussed in \cite{MacKay}. All input lines are excitatory.  
The neuron fires an output impulse if the number of stored impulses, $\Sigma$, 
is equal or higher than the threshold value, $N_{0}$. 
After that, BN clears its memory and is ready to receive fresh inputs.
That is, every input impulse either disappears contributing to a triggering event, 
or it is lost
after spending $\tau$ units of time in the neuron's internal memory.
The latter represents leakage. Here, the leakage is abrupt, while in more traditional models it is gradual.

The BN model is not general, but somewhat inspired by neurons as integrators
up to a threshold. Its name is suggested by binding of features/events in 
large-scale neuronal circuits \cite{Damasio89,Eckhorn,Engel91a}. 
Its operational simplicity is provided by the fact that each input impulse traces 
entirely disappear after finite time $\tau$. This is in the contrast to more familiar models
where the traces (excitatory postsynaptic potentials, EPSP) decay exponentially.
E. g., in the leaky integrate-and-fire model, EPSP is mimicked as pure exponential function
the traces of which can disappear completely only after triggering.
In the BN model, the EPSP is mimicked as box function of width/duration $\tau$
and the traces are stored in the neuron no longer than $\tau$ units of time.

Further, we expect that input stream in each input line is the
Poisson one with some intensity $\lambda_i$. In this case, all input
lines can be collapsed into a single one delivering Poisson stream
of intensity $\lambda=\sum_i\lambda_i$, see Figure \ref{fig:BNDF}.

For analytic derivation, we use BN with $N_0=2$ in order to keep
mathematical expressions shorter. It seems, that cases with higher
thresholds might be considered with the same approach, but even
$N_0=3$ without feedback requires additional combinatorial efforts, see \cite{Vid5}.
Therefore, cases of higher threshold are tested here only numerically.

As regards real biological neurons, 
the number of synaptic impulses in the internal memory which is necessary to trigger a neuron, 
varies from one \cite{Miles}, through fifty \cite{Barbour}, to 60-180 \cite{Andersen2}, and 100-300 \cite{Andersen}.

\subsection{Feedback line action}
\label{sec:feedback}
In real neuronal systems, a neuron can form synapses from its axonal branch to its own dendritic tree
\cite{Aron,Stevens,Chan-Palay,Sakmann,Nicoll,Park,Tamas,Loos}. Synapses of this type are called autapses. 
Some of the neurons forming autapses 
are known to be inhibitory, see \cite{Chan-Palay,Park,Tamas} for experimental evidence.
As a result, the neuron stimulates itself obtaining an inhibitory impulse
through an autapse after each firing 
with some propagation delay. We model this situation assuming that
 output impulses of BN  are fed back into BN's input with delay 
$\Delta$. This gives the inhibitory BN with delayed feedback model, Figure~\ref{fig:BNDF}.

The inhibitory action of feedback impulses is modeled in the following way. 
When the inhibitory impulse reaches BN, it annihilates all excitatory impulses already present in the BN's memory, 
similarly as the Cl-type inhibition shunts 
depolarization of excitable membrane, see \cite{Schmidt1981}.
If at the moment of inhibitory impulse arrival, the neuron is empty, then
the impulse disappears without any action, similarly as Cl-type inhibition
does not affect membrane's voltage in its resting state.
Such inhibition is "fast" in that sense, that the inhibitory impulses act instantaneously and are not remembered by neuron. This simple behavior
is approved by relatively fast kinetics of the chloride inhibitory postsynaptic currents
\cite{Borst}.

The feedback line either keeps one impulse, or keeps no impulses and cannot convey two or more impulses at the same time. 
Biological correlates supporting to an extent this assumption could be a prolonged refractory time
and/or short-term synaptic depression. The latter can have the recovery time up to 20 s \cite{Wu}.
If the feedback line is empty at the moment of firing, 
the output impulse enters the line, and after time interval equal $\Delta$ 
reaches the BN's
input. If the line already keeps one impulse at the moment of 
firing,  the just fired impulse ignores the line.

This means, that at the beginning of an output ISI
the feedback line is never empty.
In order to describe the state of the feedback line,
we introduce the stochastic variable $s$, $s\in\,]0;\Delta]$,
which gives the time to live of the impulse in the feedback line, see Fig~\ref{fig:BNDF}.
Hereinafter, we will use the values of $s$ just at the moments of output ISI 
beginnings (just after firings).

We assume, that time delay $\Delta$ of impulse in the feedback line is smaller than the BN's memory duration, $\tau$:
\begin{equation}
\label{case}
	\Delta<\tau.
\end{equation}
It allows to make analytic expressions shorter. Also, the assumption
(\ref{case}) is consistent with the case of direct feedback, 
not mediated by other neurons.
See also Part 1 of this paper, \cite{BNDiF}, this issue, for more detailed discussion 
and justification of this assumption.


\section{Statement of the problem}
\label{sec:problem}

The input stream of impulses, which drives neuronal activity is 
the Poisson stream. It is stochastic,
therefore, the output activity of our system requires probabilistic description
in spite of the fact that both the BN and the feedback line action mechanisms
are deterministic.
We treat the output stream of inhibitory BN with delayed feedback as the 
stationary process\footnote{
The stationarity of the output stream results both from the stationarity of the input 
one and from 
the absence of time-dependent parameters in the BN model, see Section~\ref{sec:BN}. 
In order to ensure stationarity, we also expect that system is considered 
after initial period sufficient to forget the initial conditions.
}.
In order to describe its statistics, we introduce the following basic functions:
\begin{itemize}
	\item the joint probability density $P(t_{m},t_{m-1},\ldots,t_{0})$ for $(m+1)$
successive output ISI durations, $t_0$ is the first one.
        \item the conditional probability density $P(t_{m}\mid t_{m-1},\ldots,t_{0})$ 
for output ISI durations; $P(t_{m}\mid t_{m-1},\ldots,t_{0})\rmd t_{m}$ gives the probability 
to obtain an output ISI of duration between $t_{m}$ and $t_{m}+\rmd t_{m}$ provided 
the previous $m$ ISIs had durations $t_{m-1},t_{m-2},\ldots,t_{0}$, respectively.
\end{itemize}

\begin{definition}
The sequence of random variables $\{t_{j}\}$, taking values in $\Omega$, is called the Markov chain of the order $n\ge0$, if 
\begin{displaymath}
	\forall_{m > n} \forall_{t_0\in\Omega}\ldots \forall_{t_m\in\Omega}\  
        P(t_{m}\mid t_{m-1},\ldots,t_{0}) 
	= P(t_{m}\mid t_{m-1},\ldots,t_{m-n}),
\end{displaymath}
and this equation does not hold for any $n'<n$ (see e.g. \cite{Doob}).
In the case of ISIs one reads $\Omega=\mathbb{R^+}$.
\end{definition}

In particular, taking $m=n+1$, we have the necessary condition 
\begin{multline}
\label{def}
	P(t_{n+1}\mid t_{n},\ldots,t_{1},t_{0}) 
	= P(t_{n+1}\mid t_{n},\ldots,t_{1}),
\\
	t_i\in\mathbb{R^+}, \qquad i=0,\ldots,n+1,
\end{multline}
required for the stochastic process $\{t_{j}\}$ of ISIs 
to be the $n$-order Markov chain.

Our purpose in this paper is to prove the following theorem.
\begin{theorem}
\label{theo}
The output ISIs stream of inhibitory BN with delayed feedback under Poisson stimulation
cannot be represented as a Markov chain of any finite order.
\end{theorem}

\section{Main calculations}
\label{sec:main}

This section with Appendices contains the required proof of Theorem \ref{theo}.
Here we give a very short sketch of the methods we use. 

In order to prove  the Theorem \ref{theo} it is necessary and
enough to prove that (\ref{def})
does not hold.
The Definition 1 includes universal quantifiers, therefore, it is enough to
prove that $P(t_{n+1}\mid t_{n},\ldots,t_{1},t_{0})$ has a property,
which explicitly depends on the $t_0$. For the excitatory neuron case,
studied in \cite{vidybida_nmark}
such a property was the Dirac $\delta$-function singularity presence
in the $P(t_{n+1}\mid t_{n},\ldots,t_{1},t_{0})$. The position of
the singularity depends explicitly on the $t_0$. Here we use the similar
method for the inhibitory neuron. In this case the 
$P(t_{n+1}\mid t_{n},\ldots,t_{1},t_{0})$ does not have a 
$\delta$-function singularity. Instead, 
$P(t_{n+1}\mid t_{n},\ldots,t_{1},t_{0})$ has a jump type discontinuity
along certain hyperplanes. Position of these hyperplanes depends 
exactly on the $t_0$,
see (\ref{discn+1}). This proves that $t_0$-dependence of 
$P(t_{n+1}\mid t_{n},\ldots,t_{1},t_{0})$ cannot be eliminated.
Again, due to the universal quantifiers presence in the Definition 1
it is enough to prove the $t_0$-dependence at a subset of variables
$t_0,\dots,t_{n+1}$, which has nonzero measure. For the general case 
of any $n$ we use such a subset, see (\ref{domain}). For the particular
cases of $n=0$ and $n=1$, we study the whole set of possible values,
see Sec. \ref{sec:cases}, even if it is not necessary for the proof.

\subsection{Proof outline}
\label{sec:outline}
We are going to show analytically, that 
the equality (\ref{def}) does not hold for any finite value of $n$.
Namely, we will derive the exact analytic expression 
for the conditional probability density $P(t_{n+1}\mid t_{n},\ldots,t_{1},t_{0})$
and show, that it depends on $t_{0}$ for any finite number $n$.

For this purpose, we denote by $s$ the time left for an impulse in the feedback
line to reach the neuron, see Fig.~\ref{fig:BNDF}. Hereafter, we call $s$
as "time to live" of the impulse in the feedback line.
From the 
Sec.~\ref{sec:feedback} it follows: the feedback line always conveys an impulse 
at the moment when an ISI starts.
This allows us to
introduce the stream $\mathbf{ts}$ of events $(t,s)$
\begin{displaymath}
	\mathbf{ts} = \{\dots, (t_i,s_i),\dots\},
\end{displaymath}
where $s_i$ is the time to live of the impulse in the feedback line at the moment, when the ISI $t_i$ starts.
We consider the joint probability density $P(t_{n+1}, s_{n+1}; t_{n},s_{n};\ldots;t_{0} ,s_{0})$ for realization of $(n+2)$ successive events $(t,s)$, and the corresponding conditional probability density $P(t_{n+1},s_{n+1}\mid t_{n},s_{n};\ldots;t_{0},s_{0})$ for these events. 

Then, we proof the following lemma,
which will be used in 
our calculations.
\begin{lemma}
\label{lemma}
Stream $\mathbf{ts}$ is the 1-st order Markovian:
\begin{multline}
\label{marka}
	\forall_{n\ge0}
	\forall_{t_{0}>0} \forall_{s_0\in\,]0;\Delta]}\ldots
	\forall_{t_{n+1}>0} \forall_{s_{n+1}\in\,]0;\Delta]}
\\ 
	P(t_{n+1},s_{n+1}\mid t_{n},s_{n};\ldots;t_{0},s_{0}) 
	= P(t_{n+1},s_{n+1}\mid t_{n},s_{n}),
\end{multline}
where $\{t_0,\ldots,t_{n+1}\}$ is the set of successive ISIs,
and $\{s_0,\ldots,s_{n+1}\}$ are the corresponding times to live.
\end{lemma}
See Appendix \ref{lemmaproof} for the proof.

Then, in order to find the conditional probability density $P(t_{n+1}\mid t_{n},\ldots,t_{1},t_{0})$, we perform the following steps:
\begin{itemize}
	\item \emph{Step 1.} Use the property (\ref{marka}) for calculating joint probability density of events $(t,s)$:
\begin{multline}
\label{mark}
	P(t_{n+1},s_{n+1};t_{n},s_{n};\ldots;t_{0},s_{0}) = 
\\
	P(t_{n+1},s_{n+1}\mid t_{n},s_{n}) 
	\ldots P(t_{1},s_{1}\mid t_{0},s_{0}) P(t_{0},s_{0}),
\end{multline}
where $P(t,s)$ and $P(t_{n},s_{n}\mid t_{n-1},s_{n-1})$ denote the stationary
probability density and conditional probability density (transition probability) 
for events $(t,s)$.

	\item {\emph {Step 2.}} Represent $P(t_{n+1},t_{n},\ldots,t_{0})$ as marginal probability
by integration over variables $s_i,\, i=0,1,\dots,n+1$:
\begin{multline}
\label{main}
	P(t_{n+1},t_{n},\ldots,t_{0}) = 
\\
	\int_{0}^{\Delta}\rmd s_{0} 
	\int_{0}^{\Delta}\rmd s_{1} \ldots \int_{0}^{\Delta}\rmd s_{n+1} 
	P(t_{n+1},s_{n+1};t_{n},s_{n};\ldots;t_{0},s_{0}).
\end{multline}

	\item \emph {Step 3.} Use the definition of conditional probability density:
\begin{equation}
\label{defcond}
	P(t_{n+1}\mid t_{n},\ldots,t_{1},t_{0})
	= \frac{P(t_{n+1},t_{n},\ldots,t_{0})}
	{P(t_{n},\ldots,t_{0})}.
\end{equation}

\end{itemize}

Taking into account the Steps 1 and 2, one derives for the joint probability density
\begin{multline}
\label{joint}
	P(t_{n+1},t_{n},\ldots,t_{0}) = 
\\
	\int_{0}^{\Delta}\rmd s_{0} \ldots \int_{0}^{\Delta} \rmd s_{n+1}
	P(t_{0},s_{0})\ 
	\prod_{k=1}^{n+1} P(t_{k},s_{k}\mid t_{k-1},s_{k-1}).
\end{multline}

In the next sections, we are going to find the exact analytic expressions for probability densities $P(t,s)$ and $P(t_k,s_k\mid t_{k-1},s_{k-1})$, 
and perform the integration in (\ref{joint}). Then we will apply the Step 3, 
above, to find expressions for the conditional probability densities 
$P(t_{n+1}\mid t_{n},\ldots,t_{0})$. It appears, that 
$P(t_{n+1}\mid t_{n},\ldots,t_{0})$ is a function with jump discontinuities. 
In order to prove that the equality~(\ref{def}) does not hold for any
$n\ge0$, we analyze the positions of those jump discontinuities only.

\subsection{Probability density $P(t,s)$ for events $(t,s)$}
\label{sec:Pts}
The probability density $P(t,s)$ can be derived as the product
\begin{equation}
\label{P(t,s)}
	P(t,s) = F(t\mid s) f(s).	
\end{equation}
Here $F(t\mid s)$ denotes conditional
probability density for ISI duration provided the time to live of the impulse in the feedback line equals $s$ at the moment of this ISI beginning. 
The exact expression for $F(t\mid s)$ is calculated in Eqs. (9)--(11) of
the first part of this paper, see \cite{BNDiF}, this issue. This is done based on the
definition of BN with delayed inhibitory feedback by considering
different relationships between $t$ and $s$. In   \cite{BNDiF}, this issue, we use
notation $P^\Delta(t\mid s)$, here we use $F(t\mid s)$ instead, in order to
make final expressions shorter. As a result we have found in \cite{BNDiF},
this issue,
the following expression
\begin{equation}
\label{P(t|s)sing}
	F(t\mid s)=
	\begin{cases}
		\lambda^2 t\ e^{-\lambda t},\quad t\in\,]0;s[,\\\\
		(1+\lambda s)\ e^{-\lambda s}P^0(t-s),\quad t\ge s,
	\end{cases}
\end{equation}
were $P^0(t)$, $t>0$, denotes an output ISI probability density for BN without feedback, which was obtained in \cite[Eq. (3)]{Vid5}.
Explicit expressions for $P^0(t)$ are different for different domains of $t$.
For example, 
\begin{equation}
\label{P0}
	P^0(t)=	\lambda^2 t\ e^{-\lambda t},\quad t\in\,]0;\tau].
\end{equation}
It is proven in \cite{Vid5}, that $P^0(t)$ is a continuous function for whole range
 of ISI durations: $t\in\,]0;\infty[$.

Another function in  (\ref{P(t,s)}), $f(s)$, denotes the stationary probability
density for time to live of the impulse in the feedback line at the moment of
an output ISI beginning. The exact expression for the $f(s)$ is found
in the first part of this paper, see Eqs. (14)--(16) in \cite{BNDiF}, this issue. 
This is done by the following method. 
First, we calculate the transition probability density, $P(s'\mid s)$,
which gives the probability to have an impulse in the feedback line with
time to live in $[s';s'+ds'[$ at the moment an ISI starts, 
provided that at the moment when the previous ISI starts, there was an impulse
in the feedback line with time to live equal $s$. The exact expression for the 
$P(s'\mid s)$, see \cite[Eq. (13)]{BNDiF}, this issue,
is found based on the exact expression (\ref{P(t|s)sing})
for the $F(t\mid s)$. Exact expression for $f(s)$
is then found as normalized solution to the following equation
$$
\int_{0}^{\Delta} P(s'\mid s)\,f(s)\,ds=f(s').
$$
We do not need the exact expression for $f(s)$ here,
(see the first part of this paper, \cite[Eq. (15)]{BNDiF}, this issue, for the exact expression).
What do we need here is the form of $f(s)$, which is 
\begin{equation}
\label{fsing}
	f(s) = a \cdot \delta(s-\Delta)+g(s),\quad\textrm{where}\quad a=\frac{4\rme^{2\ld}}{(3+2\ld)\rme^{2\ld}+1},
\end{equation}
where $\delta(\cdot)$ -- is the Dirac delta-function, 
$g(s)$ -- is a regular function, which vanishes out of interval $s\in\,]0;\Delta]$, 
the $a$ gives the probability to obtain the impulse in the feedback line 
with time to live equal $\Delta$ at the beginning of an arbitrary output ISI, 
$\lambda$ --- is the input Poisson stream intensity.

Let us explain the presence of Dirac $\delta$-function type singularity in $f(s)$.
The probability to have time to live, $s$, exactly equal
$\Delta$ at the moment of an output ISI beginning is not infinitesimally small.
Every time, when the line is free at the moment of an output ISI beginning, 
the impulse enters the line and has time to live equal $\Delta$. 
For the line to be free from impulses at the moment of triggering, 
it is enough that $t> s$ for the previous ISI. 
The set of realizations of the input Poisson process, each realization
satisfying  $t> s$, has non-zero probability $a$, see 
(\ref{fsing}),
and this gives the $\delta$-function at $s=\Delta$ in the probability density 
$f(s)$.

It is essential for further study,
that $F(t\mid s)$ considered as function of $t$ has a jump discontinuity at $t=s$.
Indeed, using (\ref{P(t|s)sing}) and (\ref{P0}), one obtains
\begin{alignat}{1}
\nonumber
	&\lim\limits_{t\to s-0} F(t\mid s) = 
	\lambda^2 s\ e^{-\lambda s}>0, \qquad s\in\,]0;\Delta],
\\\nonumber
	&\lim\limits_{t\to s+0} F(t\mid s) = 0.	  
\end{alignat}
We emphasize,
that $F(t\mid s)$ is a continuous function 
elsewhere except of the point $t=s$, where it has strictly positive jump.
The continuity of $F(t\mid s)$ at $t\in\,]0;s[$ and $t\in\,]s;\infty[$,
and its jump at $t=s$ will be used later.

The presence of jump in $F(t\mid s)$ at $t=s$ can be explained as follows. 
According to the definition of $F(t\mid s)$, the inhibitory impulse from the feedback line
arrives  $s$ seconds later than the ISI $t$ starts. After the inhibitory impulse arrival,
it is guaranteed, that the BN is empty. To trigger the BN just after that moment,
it is necessary to get two impulses from the input stream
within infinitesimally small time interval. This event has infinitesimally small probability for the Poisson process 
(as well as for any other point process).
That is why, the value of probability density $F(t\mid s)$ drops to zero 
at $t=s+0$
and $F(t\mid s)$ experiences discontinuity at $t=s$.
\begin{figure}
	\includegraphics[width=0.45\textwidth,angle=0]{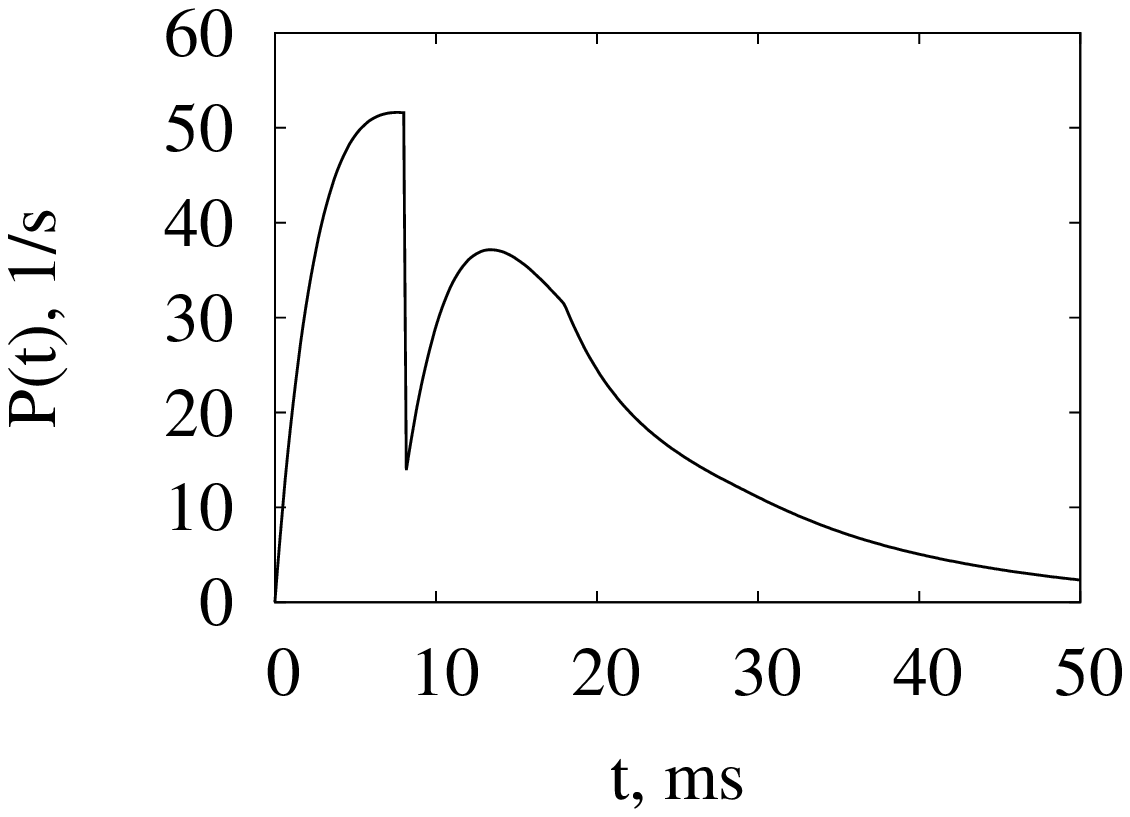}
	\includegraphics[width=0.45\textwidth,angle=0]{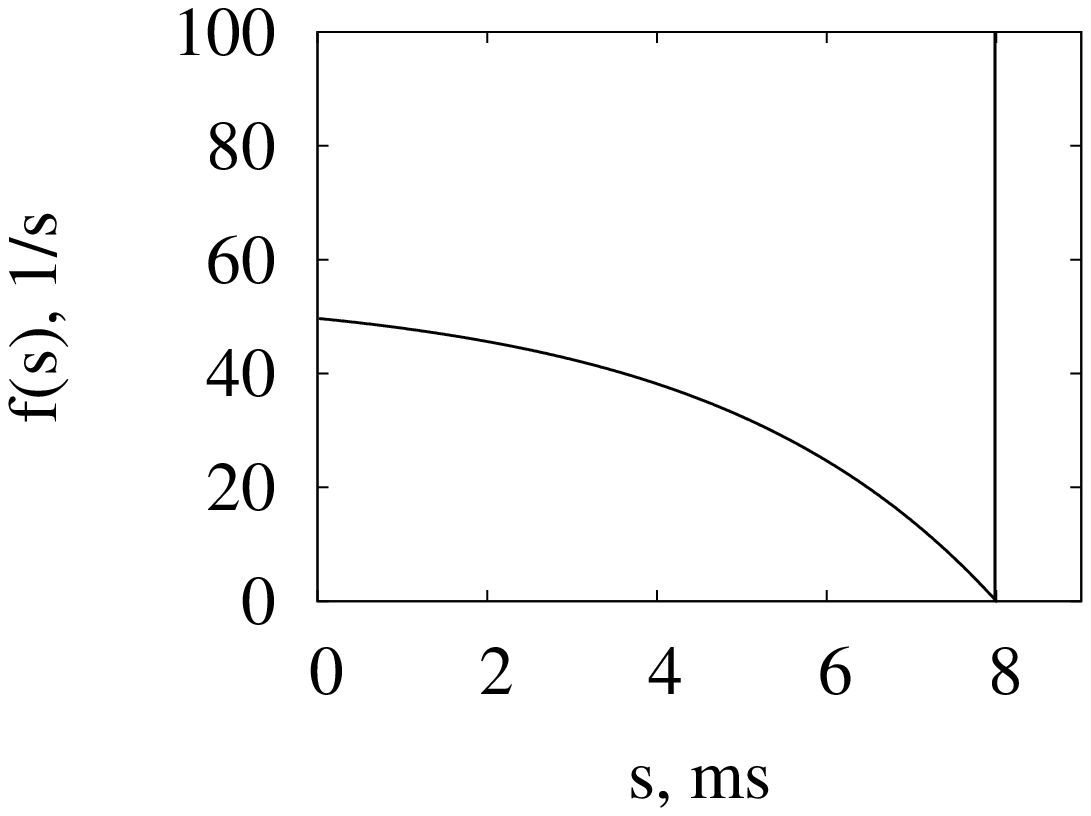}
\caption{\label{fig:Ptfs} \emph{Left}: output ISI probability density $P(t)$ reproduced from 
\cite[Fig. 2]{BNDiF}, this issue; \emph{Right}: probability density $f(s)$  
for times to live of the impulse in the feedback line.
Here $\tau$ = 10 ms, 
$\Delta$ = 8 ms, $\lambda$ = 150 s$^{-1}$, $N_0$=2. }
\end{figure}

The output ISI probability density $P(t)$ for inhibitory neuron with delayed feedback 
can be obtained as the result of integration of (\ref{P(t,s)}):
\begin{equation}
\label{P(t)}
	P(t) = \int_{0}^{\Delta} F(t\mid s)f(s)\rmd s.
\end{equation}
Discontinuity of $F(t\mid s)$ at $t=s$ and $\delta$-function type singularity at $s=\Delta$ in $f(s)$
result in discontinuity of $P(t)$ at $t=\Delta$.

Examples of $P(t)$ and $f(s)$ graphs can be found in Fig. \ref{fig:Ptfs}.

\subsection{Conditional probability density $P(t_k,s_k\mid t_{k-1},s_{k-1})$}
\label{sec:Pts|ts}
Here we find the conditional probability density 
$P(t_k,s_k\mid t_{k-1},s_{k-1})$ 
for events $(t_k,s_k)$, which determines the probability to obtain the event
$(t_k,s_k)$, with precision $\rmd t_k \rmd s_k$,
provided the previous event was $(t_{k-1},s_{k-1})$.
By definition of conditional probabilities, the probability density wanted can be represented as the following product
\begin{equation}
	P(t_k,s_k\mid t_{k-1},s_{k-1}) = 
	F(t_k\mid s_k,t_{k-1},s_{k-1}) f(s_k\mid t_{k-1},s_{k-1}),
\end{equation}
where $F(t_k\mid s_k,t_{k-1},s_{k-1})$ denotes conditional probability
density for ISI duration, $t_k$, 
provided i) this ISI started with lifetime of impulse in the feedback line
equal to $s_k$, and ii) previous $(t,s)$-event was $(t_{k-1},s_{k-1})$; 
the $f(s_k\mid t_{k-1},s_{k-1})$ denotes conditional probability density 
for times to live of impulse in the feedback line under condition ii). 
It is obvious, that
\begin{equation}
	F(t_k\mid s_k,t_{k-1},s_{k-1}) = F(t_k\mid s_k),
\end{equation}
because with $s_k$ being known, the previous event $(t_{k-1},s_{k-1})$ does not add any information, useful to predict $t_k$ (compare with the proof
of Lemma \ref{lemma}, Appendix \ref{lemmaproof}).

In order to find the probability density $f(s_k\mid t_{k-1},s_{k-1})$, let us consider various possible relations between $t_{k-1}$ and $s_{k-1}$. If $t_{k-1} \ge s_{k-1}$, the line will have time to get free from the impulse during the ISI $t_{k-1}$. That is why at the beginning of the ISI $t_k$, an output spike will enter the line and will have time to live $s_k=\Delta$ with 
 probability 1. 
 Therefore, the probability density contains the corresponding $\delta$-function:
\begin{equation}
	f(s_k\mid t_{k-1},s_{k-1}) = \delta(s_k-\Delta), \qquad t_{k-1}\ge s_{k-1}.
\end{equation}
If $t_{k-1} < s_{k-1}$, than the ISI $t_{k-1}$ ends before the impulse leaves the feedback line. Therefore, at the beginning of the $t_k$,
the line still keeps the same impulse as at the beginning of $t_{k-1}$. This impulse has time to live being equal to 
$s_k = s_{k-1}-t_{k-1}$, so
\begin{equation}
	f(s_k\mid t_{k-1},s_{k-1}) =
	\delta(s_k-s_{k-1}+t_{k-1}), \qquad t_{k-1} < s_{k-1}.	
\end{equation}
Taking all together, for the conditional probability density $P(t_k,s_k\mid t_{k-1},s_{k-1})$ one obtains
\begin{alignat}{2}
\nonumber
	P(t_k,s_k\mid t_{k-1},s_{k-1})
	&=F(t_k\mid s_k)\delta(s_k-\Delta),& \qquad&t_{k-1}\ge s_{k-1},
\\
\label{P(t,s|t,s)}
	&=F(t_k\mid s_k)\delta(s_k-s_{k-1}+t_{k-1}),& \qquad &t_{k-1} < s_{k-1},
\end{alignat}
where exact expression for $F(t\mid s)$ is given in (\ref{P(t|s)sing}).



\subsection{Joint probability density $P(t_{n+1},\ldots,t_{0})$}
\label{sec:joint}
In this section, we are going to find the exact analytic expression for the joint probability density $P(t_{n+1},\ldots,t_{0})$ 
at the following domain 
\begin{equation}
\label{domain}
	D_1=\left\{(t_0,\ldots,t_n, t_{n+1})\ \ \Big |\ \sum_{i=0}^{n}t_{i}<\Delta\right\}.
\end{equation}
Notice, that coordinate $t_{n+1}$ is not included to the condition here.
The set of $(n+2)$ 
successive ISI durations $t_0,\ldots,t_n,t_{n+1}$ has non-zero probability, 
$p_\Delta>0$, to fall into the domain (\ref{domain}). 
Indeed, BN with threshold $N_{0}=2$ requires $2(n+1)$ input impulses 
within time window $]0;\Delta[$ to be triggered $(n+1)$
 times within this window (condition 
(\ref{case}) ensures that no one input impulse will be lost). 
BN receives excitatory impulses from the Poisson stream and inhibitory impulses from the feedback line. 
But no more than one impulse from the line may have time to reach BN's input 
during time interval less than $\Delta$. Therefore, if 
 as much as $(2n+3)$ input impulses are received from the Poisson stream 
during the time interval $]0;\Delta[$, the inequality 
(\ref{domain}) holds for sure, no matter was an impulse from the feedback line involved, or not. 
Therefore,
$
	p_{\Delta} > p(2n+3,\Delta) > 0,
$
where $p(i,\Delta)$ gives the probability to obtain $i$ impulses from the Poisson stream 
during time interval $\Delta$ \cite{Feller}: 
$p(i,\Delta) = \rme^{-\ld}{(\ld)^{i}}/{i!}$. 

For a fixed $(n+2)$-tuple $(t_0,\ldots,t_n, t_{n+1})\in D_1$,
let us split the integration domain for $s_{0}$ in (\ref{joint}) in the following way: 
\begin{displaymath}
	]0;\Delta]=]0;t_0]\cup]t_0;t_0+t_1]\cup]t_0+t_1;t_0+t_1+t_2]
	\cup\cdots\cup
	]t_0+t_1+\cdots+t_n;\Delta],
\end{displaymath}
or
\begin{displaymath}
	 \int_{0}^{\Delta}\rmd s_{0}
	= \int_{0}^{t_{0}}\rmd s_{0} 
	+ \sum_{i=1}^{n}\int_{\sum_{j=0}^{i-1}t_{j}}^{\sum_{j=0}^{i}t_{j}}\rmd s_{0}
	+ \int_{\sum_{j=0}^{n}t_{j}}^{\Delta}\rmd s_{0},
\end{displaymath}
and introduce the following notations:
\begin{multline}
\label{Iidef}
	 I_{i} = \int\limits_{\sum_{j=0}^{i-1}t_{j}}^{\sum_{j=0}^{i}t_{j}}\rmd s_{0}
	\int\limits_{0}^{\Delta}\rmd s_{1} \ldots \int\limits_{0}^{\Delta} \rmd s_{n+1}
	P(t_{0},s_{0})\ 
	\prod_{k=1}^{n+1} P(t_{k},s_{k}\mid t_{k-1},s_{k-1}),
\\
	i=0,1,2,\ldots,n,
\end{multline}
\begin{equation}
\label{In+1def}
	 I_{n+1} = 
         \int\limits_{\sum\limits_{j=0}^{n}t_{j}}^{\Delta}\rmd s_{0}
	\int\limits_{0}^{\Delta}\rmd s_{1} \ldots \int\limits_{0}^{\Delta} \rmd s_{n+1}
	P(t_{0},s_{0})\ \prod_{k=1}^{n+1} P(t_{k},s_{k}\mid t_{k-1},s_{k-1}),
\end{equation}
where we assume, that $\sum_{j=j_{1}}^{j_{2}}=0$ for $j_{1}>j_{2}$.

According to (\ref{joint}), (\ref{Iidef}) and (\ref{In+1def}), 
the probability density $P(t_{n+1},\ldots,t_{0})$ 
can be obtained as 
\begin{equation}
\label{Psum}
	 P(t_{n+1},\ldots,t_{0}) = \sum\limits_{i=0}^{n+1} I_i. 
\end{equation}
Substituting $P(t_0,s_0)$ and $P(t_k,s_k\mid t_{k-1},s_{k-1})$ from expressions (\ref{P(t,s)}) and (\ref{P(t,s|t,s)})
to (\ref{Iidef}) and (\ref{In+1def}) and performing integration over variables $s_1,\ldots,s_{n+1}$,
one obtains
\begin{multline}
\label{Ii}
	 I_i = \prod\limits_{k=i+1}^{n+1} F(t_{k}\mid \Delta - \sum_{j=i+1}^{k-1} t_j)
	\int\limits_{\sum_{j=0}^{i-1}t_j}^{\sum_{j=0}^i t_j} 
	\prod\limits_{k=0}^{i} F(t_k \mid s_0 - \sum_{j=0}^{k-1}t_j) g(s_0) \rmd s_0,
\\
	\qquad i=0,1,2,\ldots,n.
\end{multline}
\begin{equation}
\label{In+1}
	 I_{n+1}
	= \int\limits_{\sum_{j=0}^{n}t_{j}}^{\Delta} 
	\prod\limits_{k=0}^{n+1}	
	F(t_{k}\mid s_{0}-\sum_{j=0}^{k-1}t_{j}) g(s_{0}) \rmd s_{0}
	+ a\ \prod\limits_{k=0}^{n+1}	
	F(t_{k}\mid \Delta-\sum_{j=0}^{k-1}t_{j}),
\end{equation}
where $F(t\mid s)$ and $g(s)$ were defined in (\ref{P(t|s)sing}) and (\ref{fsing})
(see Appendix \ref{sec:Ii} for the details of integration).

Taking into account (\ref{Psum}), (\ref{Ii}) and (\ref{In+1}), 
one obtains the following expression for the joint probability density for output ISI durations:
\begin{alignat}{1}
\nonumber
	P(t_{n+1}&,\ldots,t_{0}) =
	\sum_{i=0}^{n+1} I_{i}
\\\nonumber
	&=\sum_{i=0}^{n}  
	\prod\limits_{k=i+1}^{n+1}	
	F(t_{k}\mid \Delta-\sum\limits_{j=i+1}^{k-1} t_{j}) \nonumber
	\int\limits_{\sum_{j=0}^{i-1}t_{j}}^{\sum_{j=0}^{i}t_{j}}
	g(s_{0}) \prod\limits_{k=0}^{i}	
	F(t_{k}\mid s_0-\sum\limits_{j=0}^{k-1} t_{j}) \rmd s_{0} 
\\\nonumber
	&+
	\int\limits_{\sum\limits_{j=0}^{n}t_{j}}^{\Delta} 
	g(s_{0}) \prod\limits_{k=0}^{n+1}F(t_{k}\mid s_{0}-\sum\limits_{j=0}^{k-1} t_{j})
	\rmd s_{0}
	+ a \prod\limits_{k=0}^{n+1}F(t_{k}\mid \Delta-\sum\limits_{j=0}^{k-1} t_{j}),
\\
\label{P}
	&\qquad\qquad\qquad\qquad\qquad\qquad\qquad\quad  \sum_{i=0}^{n}t_{i}<\Delta,\qquad n=0,1,...,
\end{alignat}
where we assume, that $\sum_{j=j_{1}}^{j_{2}}=0$ and $\prod_{j=j_{1}}^{j_{2}}=1$ for $j_{1}>j_{2}$.

The expression 
(\ref{P}) gives the joint probability density $P(t_{n+1},\ldots,t_{0})$ 
for consecutive ISI durations at the domain $D_1$ for an arbitrary $n$. 
Therefore, the conditional probability density $P(t_{n+1}\mid t_n,\ldots,t_{0})$ 
at $D_1$ can be obtained readily, see equation  
(\ref{defcond}).  



\subsection{Discontinuities in $P(t_{n+1},\ldots,t_{0})$}
\label{sec:joint_sing}
In this section, we will answer two following questions:
i) does the $P(t_{n+1},\ldots,t_{0})$ contain 
discontinuities at $D_1$? and
ii) if it does, what are the positions of that discontinuities?

In order to ascertain the continuity
of expression, 
defined in (\ref{P}), let us first 
analyze the behavior of $I_{i}$, $i=0,\ldots,n$, and $I_{n+1}$ separately. 

Consider $I_i$, defined in (\ref{Ii}). 
Since, at $D_1$, $t_{k} < \Delta-\sum_{j=i+1}^{k-1}t_{j}$ for any $k=i+1,\ldots,n$,
the functions $F(t_{k}\mid \Delta-\sum_{j=i+1}^{k-1}t_{j})$ 
 are continuous, see (\ref{P(t|s)sing}). 
The factor $F(t_{n+1}\mid \Delta - \sum_{j=i+1}^{n}t_{j})$
undergoes a nonzero jump discontinuity when point $(t_0,\ldots, t_{n+1})$ transverses the
hyperplane defined as
\begin{equation}
\label{disci}
	 \sum\limits_{j=i+1}^{n+1}t_{j} = \Delta,
	\qquad i=0,\ldots,n,
\end{equation}
and is continuous function anywhere else.
The result of integration in (\ref{Ii}) is a continuous function
in $D_1$, see the proof in Appendix \ref{continuity}.
Therefore, at the domain $D_1$, each $I_i$ has a discontinuity of a jump type
at the hyperplane defined in (\ref{disci}).

Now, consider the continuity of $I_{n+1}$, expression 
(\ref{In+1}).
The first term, again, is a continuous function in $D_1$, the proof
is similar to what is done in  Appendix \ref{continuity}.
The only discontinuity in the second term at the domain $D_1$
is due to the factor $F(t_{n+1}\mid\Delta-\sum_{j=0}^{n}t_j)$
and it is located at the hyperplane defined as
\begin{equation}
\label{discn+1}
	 \sum\limits_{j=0}^{n+1}t_{j} = \Delta,
\end{equation}
while all $F(t_{k}\mid \Delta-\sum_{j=0}^{k-1}t_{j})$,  $k=0,\ldots,n$ are continuous functions at this domain, see (\ref{P(t|s)sing}).

According to (\ref{Psum}), 
the probability density $P(t_{n+1},\ldots,t_{0})$ 
can be obtained as a sum of all $I_i$, $i=0,\ldots,n$ and $I_{n+1}$.
Therefore, it inherits all the discontinuities, contained in $I_i$ and $I_{n+1}$.
So, at the domain $D_1$, the probability density $P(t_{n+1},\ldots,t_{0})$ 
has nonzero jump discontinuities at the $(n+2)$ hyperplanes\footnote{Note, that all hyperplanes, defined in
 (\ref{disci}) and (\ref{discn+1}) are different.}
defined in (\ref{disci}) and (\ref{discn+1}), and is a continuous function at the rest of the domain.


\subsection{Discontinuities in $P(t_{n+1}\mid t_n,\ldots,t_{0})$}
\label{sec:cond_sing}

Conditional probability density $P(t_{n+1}\mid t_{n},\ldots,t_{0})$
can be easily derived from (\ref{P}) according to the definition (\ref{defcond}).
It should be outlined, that joint probability density $P(t_{n},\ldots,t_{0})$
is strictly positive for any $(n+1)$-tuple of positive values  $(t_{n},\ldots,t_{0})$
as it can be concluded from (\ref{P}). Moreover, $P(t_{n},\ldots,t_{0})$ is
continuous at the domain
\begin{equation}\label{D2}
\sum_{i=0}^{n} t_{i} < \Delta.
\end{equation}
Indeed, at the domain (\ref{D2}), we have also
$ 
\sum_{i=0}^{n-1} t_{i} < \Delta,
$
which means that the discontinuities of $P(t_{n},\ldots,t_{0})$ are located at 
hyperplanes defined by  conditions (\ref{disci}) and (\ref{discn+1}) with $(n-1)$ substituted instead of $n$. But those conditions are never satisfied due to (\ref{D2}).
Thus, division of $P(t_{n+1},\ldots,t_{0})$ by strictly positive and continuous 
function $P(t_{n},\ldots,t_{0})$ neither does
add new discontinuities, nor does it eliminate already found in the $P(t_{n+1},\ldots,t_{0})$
at the domain $D_1$.

Therefore, at the domain $D_1$, function $P(t_{n+1}\mid t_{n},\ldots,t_{0})$ 
contains $(n+2)$ jump discontinuities, located at the same positions as in $P(t_{n+1},\ldots,t_{0})$,
equations  (\ref{disci}) and (\ref{discn+1}),
and is a continuous function at the rest of $D_1$.
The location of discontinuity (\ref{discn+1}) depends on $t_0$. 
This dependence 
cannot be compensated by any summands,  continuous at hyperplane (\ref{discn+1}),
therefore, the whole conditional probability density $P(t_{n+1}\mid t_{n},\ldots,t_{0})$ depends on $t_{0}$. 
This means,  that the condition (\ref{def}) does not hold for any $n$ for the output stream of BN with delayed feedback. 
The Theorem 1 
is proven.
\qed



\section{Particular cases}
\label{sec:cases}
In the previous sections, we have proven
the impossibility to represent the stream of output ISI durations for BN with delayed feedback as a 
Markov chain of any finite order. In particular, output ISI stream is neither a sequence of independent random variables, and therefore is non-renewal, nor 
it is the first-order Markovian process.

In the course of proving Theorem 1,
we have obtained the expression for $P(t_{n+1}, t_{n},\ldots,t_{0})$ at the domain $\sum_{i=0}^{n}t_{i}<\Delta$ in general case of an arbitrary $n$,
see 
(\ref{P}).
This allows to calculate the conditional probability density $P(t_{n+1}\mid t_{n},\ldots,t_{0})$ for $\sum_{i=0}^{n}t_{i}<\Delta$
and $n=0,1,\ldots$.

In this section, we consider two particular cases of $P(t_{n+1}\mid t_{n},\ldots,t_{0})$ when $n=0$ and $n=1$, namely, the single-ISI conditional probability density $P(t_{1}\mid t_{0})$ and the double-ISI conditional probability density $P(t_{2}\mid t_{1},t_{0})$ and obtain the expressions for $P(t_{1}\mid t_{0})$ and $P(t_{2}\mid t_{1},t_{0})$ for domain (\ref{domain}), as well 
as for all other possible domains, which were omitted in calculations with arbitrary $n$.

\subsection{Conditional probability density $P(t_{1}\mid t_{0})$}
\label{sec:corr}

In order to derive the exact expression for conditional probability density $P(t_{1}\mid t_{0})$ 
for neighbouring ISI durations, we take Steps 1--3, outlined in Section~\ref{sec:outline}, for $n=0$. 
In the case of $P(t_{1}\mid t_{0})$, there are only two domains, 
on which the expressions should be obtained separately, namely cases $t_{0}<\Delta$ and $t_{0}\ge \Delta$. 
Performing integration in (\ref{joint}), one obtains the following expressions for $P(t_{1},t_{0})$ at these domains:
\begin{alignat}{2}
\nonumber
	P(t_{1},t_{0}) &= F(t_{1}\mid \Delta)P(t_{0}),&
	\qquad& t_{0}\ge\Delta,
\\
\nonumber
	&= F(t_{1}\mid \Delta) \int\limits_{0}^{t_{0}}F(t_{0}\mid s_{0})g(s_{0})\rmd s_{0}&&
\\
\label{P2joint>}	
	&+ \int\limits_{t_{0}}^{\Delta}F(t_{1}\mid s_{0}-t_{0}) F(t_{0}\mid s_{0})f(s_{0})\rmd s_{0},&
	\qquad	& t_{0}<\Delta.
\end{alignat}

Expressions (\ref{P2joint>}) can be understood as follows. 
Since $t_{0}\ge\Delta$, one can be sure that the line has time to get free from impulse during $t_{0}$, 
therefore at the moment of next firing (at the beginning of $t_{1}$) the impulse enters the line 
and has time to live equal $\Delta$. In the case of $t_{0}<\Delta$, see (\ref{P2joint>}), two possibilities arise. 
The first term corresponds to the scenario, when the feedback line discharges conveyed impulse within time interval $t_{0}$, 
and the second one represents the case when at the beginning of $t_{1}$ the line still keeps the same impulse 
as at the beginning of $t_{0}$.

Then, using (\ref{defcond}) and (\ref{fsing}), one obtains:
\begin{alignat}{2}
\nonumber
	P(t_{1}\mid t_{0}) &= F(t_{1}\mid \Delta),&
	\qquad& t_{0}\ge\Delta,
\\
\nonumber
	&= 
	\frac{1}{P(t_{0})}\ \Big(
	F(t_{1}\mid \Delta) \int\limits_{0}^{t_{0}}F(t_{0}\mid s_{0})g(s_{0})\rmd s_{0}
	+ & a F(t_{1}&\mid \Delta -t_{0}) F(t_{0}\mid \Delta)
\\
\label{Ptt}
	&+ \int\limits_{t_{0}}^{\Delta}F(t_{1}\mid s_{0}-t_{0}) F(t_{0}\mid s_{0})g(s_{0})\rmd s_{0}\Big),&
	\qquad  &t_{0}<\Delta.
\end{alignat}
It should be outlined, that the output ISI probability density $P(t_{0})$ 
is strictly positive and continuous function at the domain $0< t_{0}<\Delta$. 
Indeed, due to (\ref{P(t|s)sing})--(\ref{P(t)}), the only discontinuity 
contained in $P(t_0)$ is placed at $t_0=\Delta$, see Figure~{\ref{fig:Ptfs}}~(a).

It can be shown, that the following normalization conditions take place:\\
$\int\limits_{0}^{\infty}\rmd t_1 P(t_1\mid t_0)=1$, and
$\int\limits_{0}^{\infty}\rmd t_0 P(t_1, t_0)=P(t_1)$.

Using (\ref{P(t|s)sing}) and (\ref{Ptt}),
one obtains the positions of
discontinuities in $P(t_{1}\mid t_{0})$:
\begin{alignat}{2}
\label{Pttsing1}
	&t_{1}=\Delta,&	\qquad &\text{if } t_{0}\ge\Delta,
\\
\label{Pttsing2}
	&t_{1}=\Delta, \qquad t_{0}+t_{1}=\Delta,& \qquad  &\text{if }t_{0}<\Delta.
\end{alignat}
Obviously, expressions 
(\ref{Pttsing2}) could be obtained directly from 
(\ref{disci}) and (\ref{discn+1}) by substituting $n=0$.

\begin{figure}
        \includegraphics[width=0.5\textwidth,angle=0]{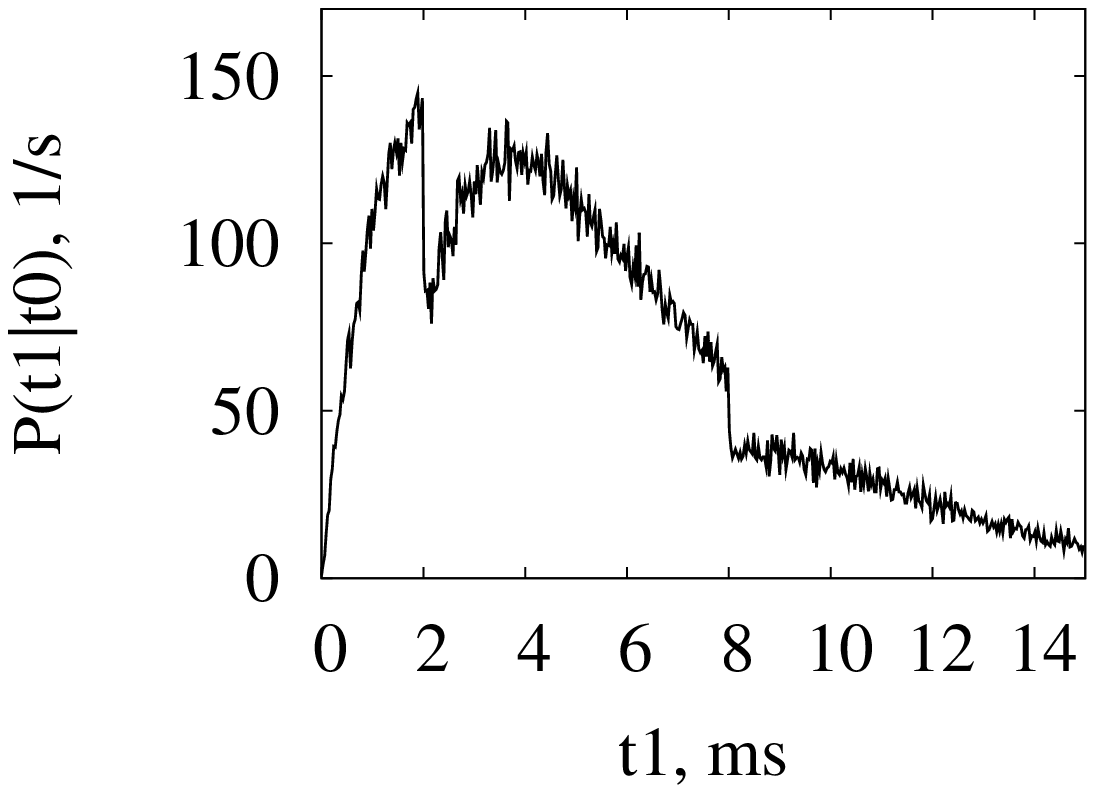}
        \includegraphics[width=0.5\textwidth,angle=0]{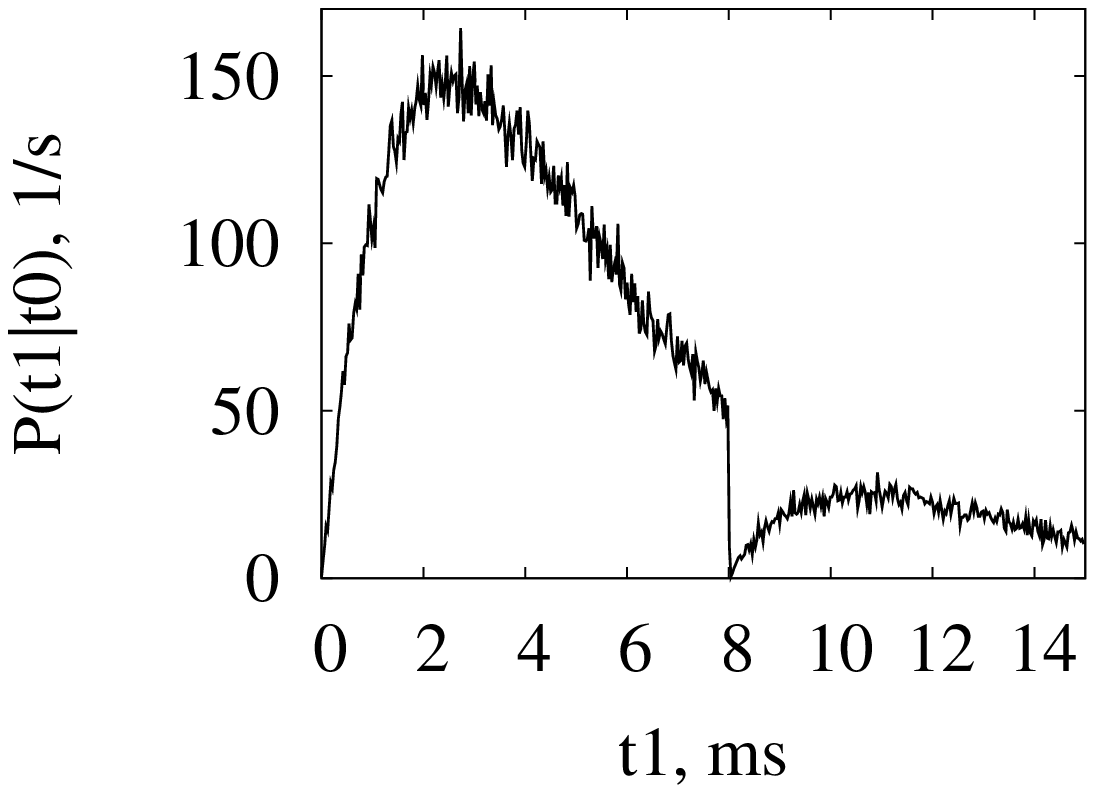}
\caption{Conditional probability density $P(t_{1}\mid t_{0})$ for $\tau$ =
10 ms, $\Delta$ = 8 ms, $\lambda$ = 400 s$^{-1}$, $N_0=2$, $t_{0}$=6 ms
(left) and $t_{0}$= 11 ms (right), found numerically by means of
Monte-Carlo method (the number of firings accounted $N=150\,000$).}
\label{fig:Ptt}
\end{figure}

As it can be seen from (\ref{Pttsing1}) and (\ref{Pttsing2}), 
the number of jump discontinuities  in $P(t_{1}\mid t_{0})$ 
and their positions  depend on $t_{0}$. 
Therefore, the conditional probability density $P(t_{1}\mid t_{0})$ 
cannot be reduced to output ISI probability density $P(t_{1})$. 
Therefore, the neighbouring output ISIs of BN with delayed feedback are correlated, as expected.

Examples of $P(t_{1}\mid t_{0})$, found for two domains numerically, by
means of Monte-Carlo method (see Section~\ref{sec:num} for details),
 are placed at Figure~\ref{fig:Ptt}.


\subsection{Conditional probability density $P(t_{2}\mid t_{1},t_{0})$}
\label{sec:mark}
In order to derive the exact expression for conditional probability density $P(t_{2}\mid t_{1},t_{0})$
 for the successive ISI durations, we take Steps 1--3, outlined in 
Section~\ref{sec:outline}, for $n=1$. In the case of $P(t_{2},t_{1},t_{0})$, there are 
five domains, on which the expressions should be obtained separately, namely,
the domain
\begin{displaymath}
	D_1=\{(t_0, t_1,t_2)\mid t_1+t_0<\Delta\},
\end{displaymath}
which was already utilized in Section~\ref{sec:main}, and the four remaining: 
\begin{alignat}{1}
\nonumber
	D_2&=\{(t_0, t_1,t_2)\mid\quad t_0\ge\Delta \quad\textrm{and}\quad t_1\ge\Delta\},
\\\nonumber
	D_3&=\{(t_0, t_1,t_2)\mid\quad t_0<\Delta \quad\textrm{and}\quad t_1\ge\Delta\},
\\\nonumber
	D_4&=\{(t_0, t_1,t_2)\mid\quad t_0\ge\Delta \quad\textrm{and}\quad t_1<\Delta\},
\\\nonumber
	D_5&=\{(t_0, t_1,t_2)\mid\quad t_0<\Delta \quad\textrm{and}\quad \Delta-t_0 \le t_1 < \Delta\},
\end{alignat}

Expressions for $P(t_2\mid t_1,t_0)$ can be found exactly on each domain:
\begin{alignat}{2}
\nonumber
	P(t_{2}\mid t_{1},t_{0}) &= F(t_{2}\mid \Delta),&
	\qquad	&(t_0, t_1, t_2)\in D_2,
\\
\nonumber
	&= F(t_{2}\mid \Delta)&
	\qquad	&(t_0, t_1, t_2) \in D_3,
\\
\nonumber
	&= F(t_{2}\mid \Delta-t_{1}),&
	\qquad	&(t_0, t_1, t_2) \in D_4,
\end{alignat}
\begin{alignat}{1}
\nonumber
	&= \frac{1}{P(t_{1},t_{0})}\ \Big(
	F(t_{2}\mid \Delta-t_{1}) F(t_{1}| \Delta) 
	\int_{0}^{t_{0}}F(t_{0}\mid s_{0})g(s_{0})\rmd s_{0}
\\
\nonumber
	&+ F(t_{2}| \Delta)
	\int_{t_{0}}^{\Delta}F(t_{1}| s_{0}-t_{0})F(t_{0}| s_{0})g(s_{0})\rmd s_{0}
	+ a\ F(t_{2}| \Delta)
	F(t_{1}| \Delta-t_{0})F(t_{0}| \Delta)\Big),
\\
\nonumber
	&\qquad\qquad\qquad\qquad\qquad\qquad\qquad\qquad\quad\quad\quad\quad\qquad\quad\quad	
	\ \ (t_0, t_1, t_2) \in D_5,
\\
\nonumber
	&= \frac{1}{P(t_{1},t_{0})}\ \Big(
	F(t_{2}\mid \Delta-t_{1}) F(t_{1}\mid \Delta) 
	\int_{0}^{t_{0}}F(t_{0}\mid s_{0})g(s_{0})\rmd s_{0}
\\
\nonumber
	&+ F(t_{2}\mid \Delta)\int_{t_{0}}^{t_{0}+t_{1}}
	F(t_{1}\mid s_{0}-t_{0})F(t_{0}\mid s_{0})g(s_{0})\rmd s_{0}
\\
\nonumber
	&+\int_{t_{0}+t_{1}}^{\Delta} F(t_{2}| s_{0}-t_{0}-t_{1})
	F(t_{1}| s_{0}-t_{0})F(t_{0}| s_{0})g(s_{0})\rmd s_{0}
\\
\label{PtttD1}
	&+ a\ F(t_{2}| \Delta-t_{0}-t_{1})
	F(t_{1}| \Delta-t_{0})F(t_{0}| \Delta)\Big),
	\quad\quad\quad\quad\quad
	(t_0, t_1, t_2) \in D_1.
\end{alignat}
where $P(t_{1},t_{0})=F(t_{1}\mid \Delta) \int_{0}^{t_{0}}F(t_{0}\mid s_{0})g(s_{0})\rmd s_{0} + \int_{t_{0}}^{\Delta}F(t_{1}\mid s_{0}-t_{0})F(t_{0}\mid s_{0})f(s_{0})\rmd s_{0}$, according to (\ref{Ptt}).

\begin{figure}
	\includegraphics[width=0.5\textwidth,angle=0]{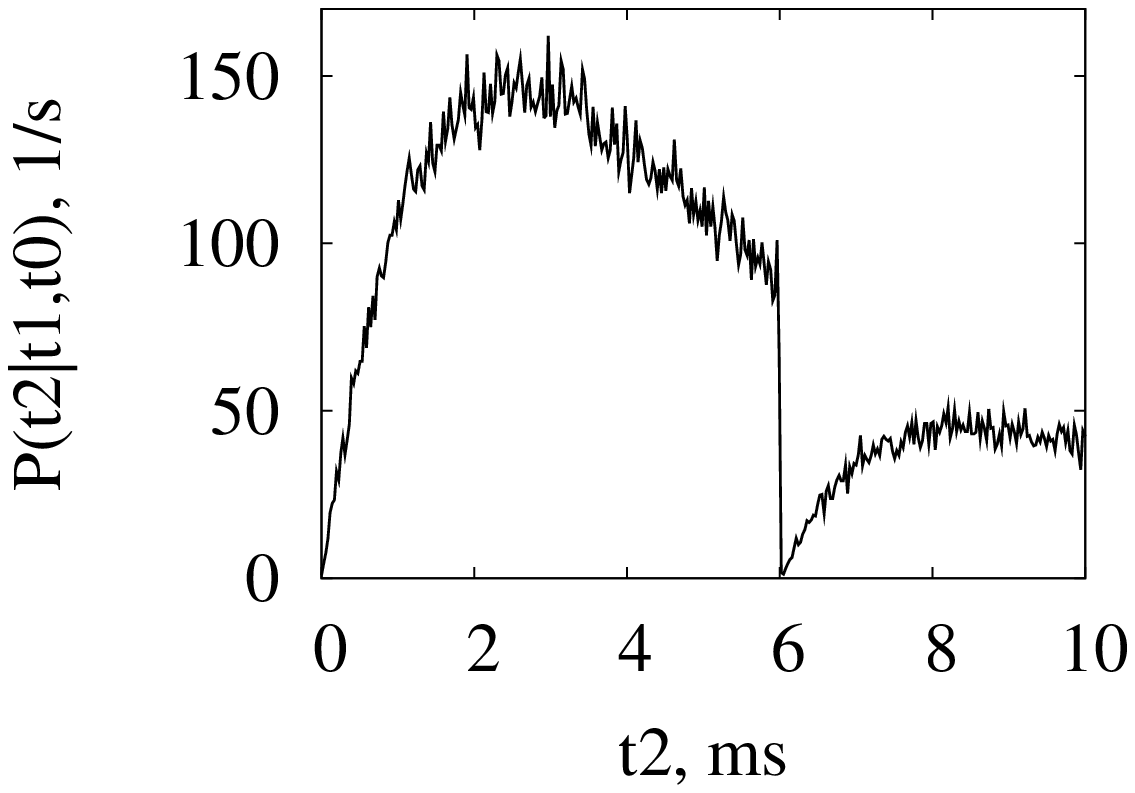}
	\includegraphics[width=0.5\textwidth,angle=0]{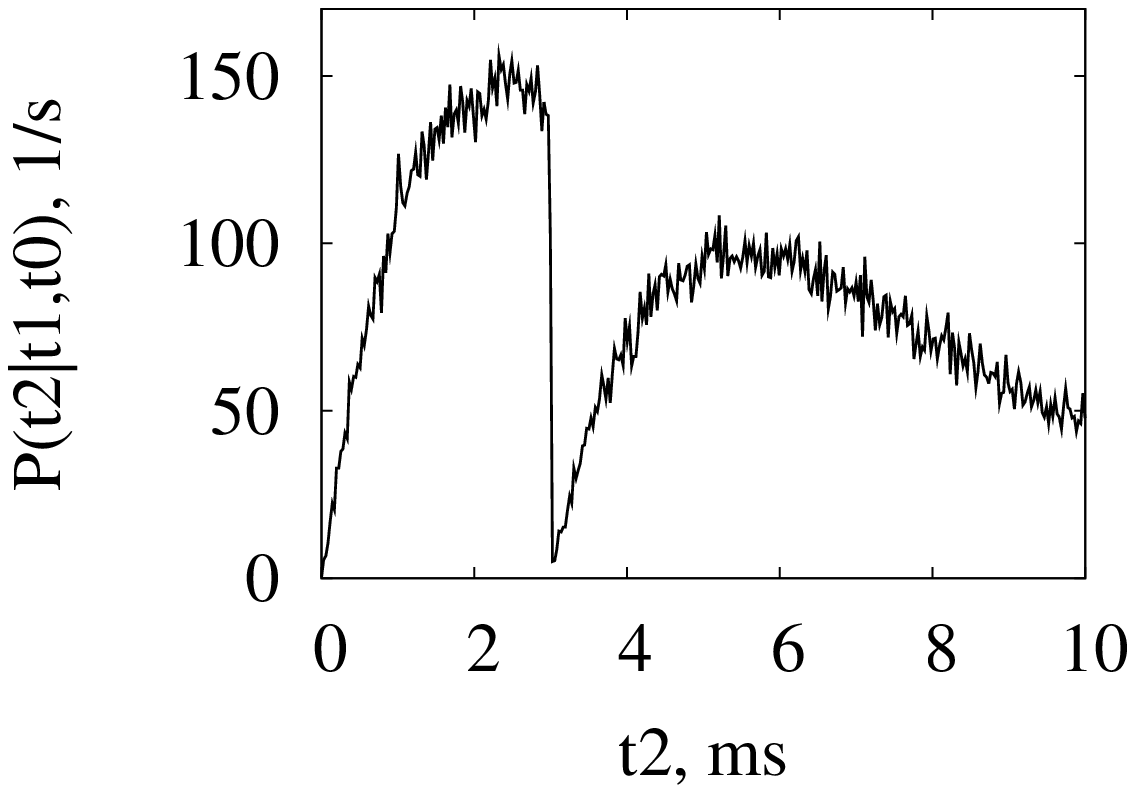}
\caption{Conditional probability density $P(t_{2}\mid t_{1}, t_0)$ 
for $\tau$ = 10 ms, $\Delta$ = 6 ms, $\lambda$ = 400 s$^{-1}$, $N_0=2$, 
$t_{1}$=8 ms, $t_{0}$=8 ms (left) and $t_{1}$ = 3 ms, $t_{0}$ = 8 ms (right), found numerically by means of Monte-Carlo method ($N=150\,000$).}
\label{fig:Pttt1}       
\end{figure}

It is worth to notice, that $P(t_{1},t_{0})$ is strictly positive and continuous function on both $D_{1}$ and $D_{5}$, 
see denominators in (\ref{PtttD1}).
Indeed, from (\ref{Pttsing1}) and (\ref{Pttsing2}) one can see, that $P(t_{1},t_{0})$ 
may include discontinuities only at the points $t_{1}=\Delta$ and $t_{1}=\Delta-t_{0}$. 
None of these points fall into $D_{1}$, or $D_{5}$.

It can be shown, that the following normalization conditions take place:\\
$\int\limits_{0}^{\infty}\rmd t_2 P(t_2\mid t_1,t_0)=1$, and
$\int\limits_{0}^{\infty}\rmd t_0 P(t_0, t_1, t_2)=P(t_2,t_1)$.

Using (\ref{P(t|s)sing}) and (\ref{PtttD1}), 
one derives 
the positions of jump discontinuities in the conditional probability density $P(t_{2}\mid t_{1}, t_0)$:
\begin{alignat}{2}
\label{Ptttsing1}
	& t_{2}=\Delta,& \qquad	&(t_0, t_1, t_2) \in D_2\cup D_3,
\\
	& t_{1}+t_{2} = \Delta,& \qquad	&(t_0, t_1, t_2) \in D_4.
\\
	& t_{2} =\Delta, \quad t_{1}+t_{2} =\Delta & \quad&(t_0, t_1, t_2) \in D_5,
\\
\label{Ptttsing5}
	& t_{2} = \Delta, \quad t_{1}+t_{2} = \Delta, \quad t_{0}+t_{1}+t_{2} = \Delta, &\quad	&(t_0, t_1, t_2) \in D_1.
\end{alignat}
Obviously, expression 
(\ref{Ptttsing5}) could be obtained directly from 
(\ref{disci}) and
(\ref{discn+1}) by substituting $n=1$.

\begin{figure}
	\includegraphics[width=0.5\textwidth,angle=0]{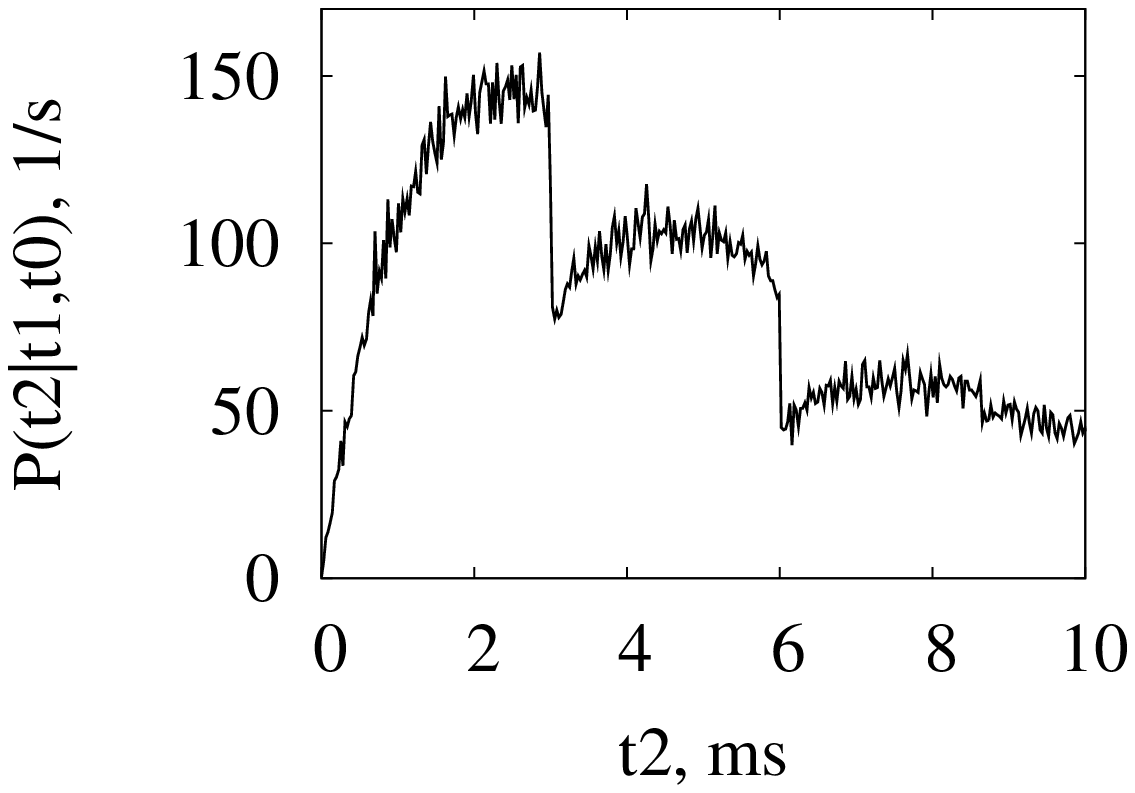}
	\includegraphics[width=0.5\textwidth,angle=0]{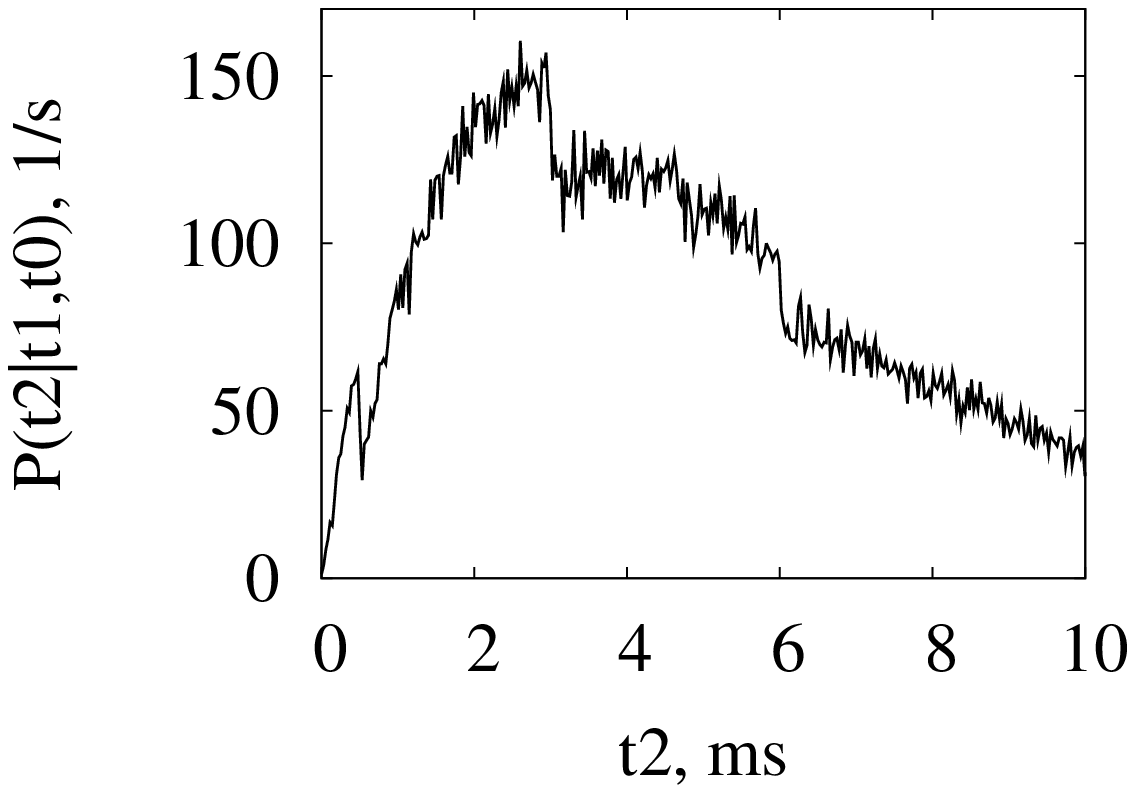}
\caption{Conditional probability density $P(t_{2}\mid t_{1}, t_0)$ for 
$\tau$ = 10 ms, $\Delta$ = 6 ms, $\lambda$ = 400 s$^{-1}$, $N_0=2$, 
$t_{1}$=3 ms, $t_{0}$=3.5 ms (left) and $t_{1}$ = 3 ms, $t_{0}$ = 2.5 ms (right), 
found numerically by means of Monte-Carlo method ($N=150\,000$).}
\label{fig:Pttt2}       
\end{figure}

As one can see, the number and the position of jump discontinuities in $P(t_{2}\mid t_{1},t_{0})$ depends on $t_{0}$, 
therefore $P(t_{2}\mid t_{1},t_{0})$ cannot be reduced to $P(t_{2}\mid t_{1})$, 
which means that the output stream is not first-order Markovian.

Examples of $P(t_{2}\mid t_{1},t_0)$, found numerically for different domains, 
are placed at Figures~\ref{fig:Pttt1} and \ref{fig:Pttt2}.



\section{Numerical simulation}
\label{sec:num}

In order to check the correctness of obtained analytic expressions,
and also
to investigate whether the output ISIs stream is non-Markovian
for inhibitory BN with higher thresholds as well as for $N_0=2$,
numerical simulations were performed. A C++ program, containing class, 
which models the operation manner of inhibitory BN with delayed feedback, was developed. 
Object of this class receives the sequence of pseudorandom numbers with Poisson probability density to its
input. The required sequences were generated by means of utilities from the GNU 
Scientific Library\footnote{http://www.gnu.org/software/gsl/}
with the Mersenne Twister generator as source of pseudorandom numbers.

\begin{figure}
	\includegraphics[width=0.5\textwidth,angle=0]{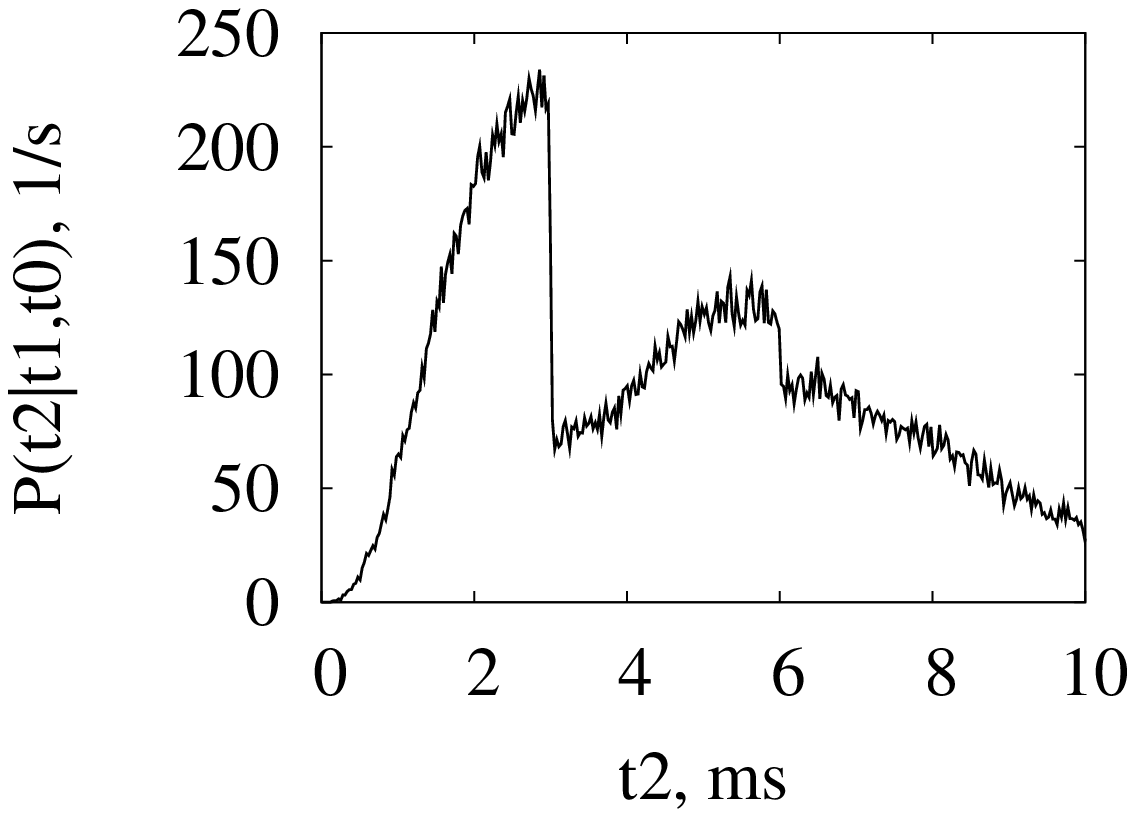}
	\includegraphics[width=0.5\textwidth,angle=0]{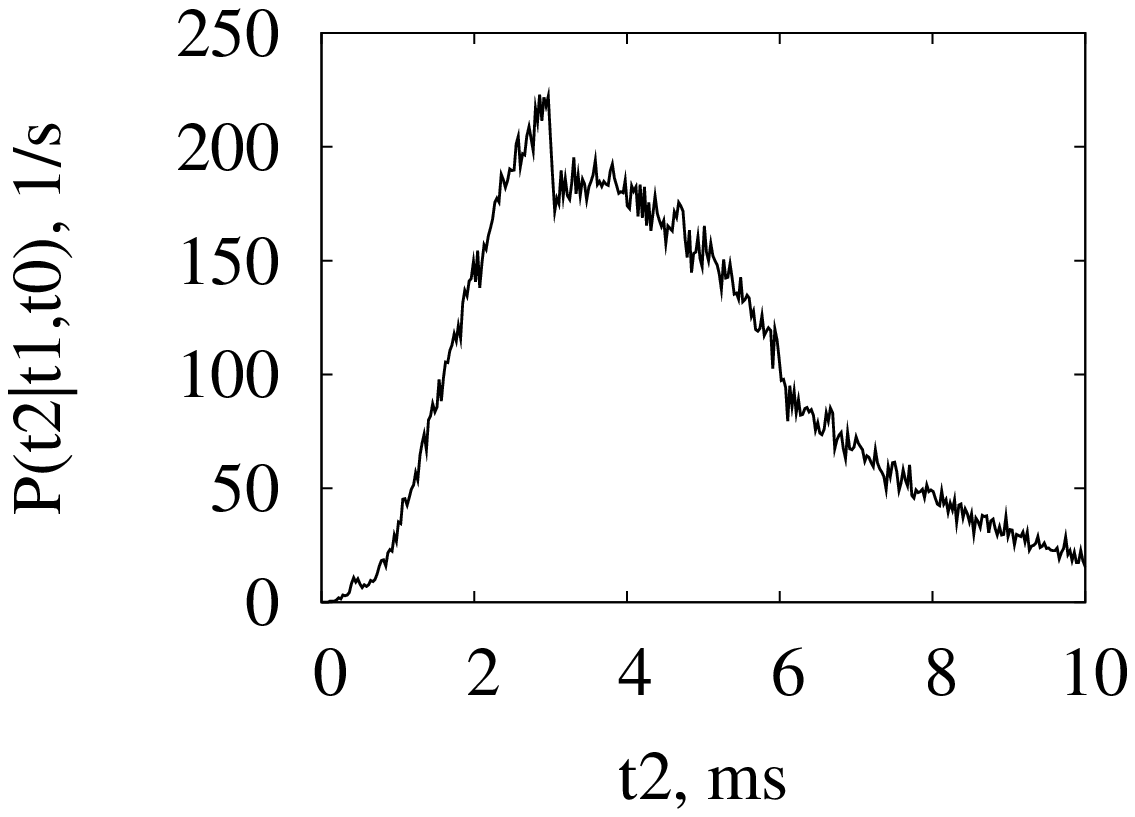}
\caption{Conditional probability density $P(t_{2}\mid t_{1}, t_0)$ for $\tau$ = 10 ms, $\Delta$ = 6 ms, 
$\lambda$ = 1000 s$^{-1}$, $N_0=4$, $t_{1}$=3 ms, $t_{0}$=3.5 ms (a) and $t_{1}$ = 3 ms, $t_{0}$ = 2.5 ms (b), 
found numerically by means of Monte-Carlo method ($N=150\,000$).
}
\label{fig:PtttN04}
\end{figure}

Program contains function, the time engine, 
which brings system to the moment just before the next input signal, 
bypassing moments, when neither external Poisson impulse, 
nor impulse from the feedback line comes. 
So, only the essential events are accounted. 
It allows one to make exact calculations faster as compared to
 the algorithm where time advances gradually by adding small time-steps.

The conditional probability densities, $P(t_{1}\mid t_{0})$
and $P(t_{2}\mid t_{1},t_{0})$, are found by counting the number 
of output ISI of different durations and normalization (see Figures~\ref{fig:Ptt} -- \ref{fig:PtttN04}). 
Obviously, for calculation of conditional distributions only those ISIs are selected, 
which follow one or two ISIs of fixed duration,  $t_{0}$ for $P(t_{1}\mid t_{0})$ 
and $\{t_{1}, t_{0}\}$ for $P(t_{2}\mid t_{1},t_{0})$. The number and the positions
of discontinuities, obtained in numerical experiments for inhibitory BN with threshold 2,
coincide with those predicted analytically in (\ref{Pttsing1}), (\ref{Pttsing2}) and (\ref{Ptttsing1}) -- (\ref{Ptttsing5}).

For $N_0>2$, conditional probability densities $P(t_{1}\mid t_{0})$
and $P(t_{2}\mid t_{1},t_{0})$ are similar to those, found for $N_0$=2.
In particular, both the quantity and position of discontinuities
coincide with those obtained for inhibitory BN with threshold 2,
as expected, 
compare Figures~\ref{fig:PtttN04}
and \ref{fig:Pttt2}.


\section{Conclusions and discussion}
\label{sec:disc}

Our results reveal the influence of delayed feedback presence on the neuronal firing statistics.
In the contrast to the cases of BN without feedback \cite{VidBiosys} and BN with instantaneous feedback~\cite{BNF}, 
the neighbouring output ISIs of inhibitory BN with delayed feedback are mutually correlated. 
This means that even in the simplest
possible recurrent network the output ISI stream cannot be treated as a renewal one. 

The non-renewalness of experimentally registered spike trains was observed
for neuronal activity in various CNS areas in mammals \cite{Lowen,Farkhooi,Nawrot07} and fish
\cite{Levine,RatnamNelson}. The simplest stochastic processes which are not
renewal are the Markov processes of various order. The order of underlying
Markov process was estimated in \cite{RatnamNelson} for activity in the 
weakly electric fish
electrosensory system. It was found in \cite{RatnamNelson} that for some neural
fibers the Markov order should be at list seven, which does not exclude that
the genuine order is higher, or that
the activity is non-Markovian. 

Actually, for proving based on experimental data
that a stochastic activity has Markov order $m$, one needs increasing amount
of data with increasing $m$. If so, it seems impossible to prove experimentally
that a stochastic activity is non-Markovian. Similarly as it is impossible
to prove experimentally that a number is irrational. 
We prove here that the output ISI stream of inhibitory BN with delayed feedback 
is non-Markovian based on complete knowledge of the mechanism which
generates the output stream. In a sense, to have this knowledge is equivalent
as to have an unlimited amount of experimental data. 

It is worth to notice, that the activity of excitatory BN with delayed feedback
is non-Markovian as well \cite{vidybida_nmark}.
We conclude, that it is namely the delayed feedback presence,
which results in non-Markovian statistics of neuronal firing.
One should take this facts into account during analysis of neuronal spike trains 
obtained from any recurrent network.


\begin{appendix}
\section{Proof of Lemma \ref{lemma}}
\label{lemmaproof}

In the compound event $(t_{n+1},s_{n+1})$, 
the time to live $s_{n+1}$ always gets its value before than the $t_{n+1}$ does.
The value of $s_{n+1}$ can be determined unambiguously from
the $(t_n,s_n)$ value (See Sections~\ref{sec:feedback} and \ref{sec:Pts|ts}):
\begin{alignat}{2}
\nonumber
		s_{n+1} &=  s_n - t_n,& \qquad &t_n < s_n, 
\\\nonumber
		&=\Delta,& \qquad &t_n \ge s_n.
\end{alignat}
	
The only two factors, which determine the next ISI duration, $t_{n+1}$, are 
(i) the value of $s_{n+1}$, and (ii) the behavior of the input Poisson
stream under the condition $(t_{n},s_{n};\ldots;t_{0},s_{0})$
after the moment $\theta$, when the  $t_{n+1}$ starts.
The $s_{n+1}$ value does not depend on $(t_{n-1},s_{n-1};\ldots;t_{0},s_{0})$, see above.
As regards the input Poisson stream, condition $(t_{n},s_{n};\ldots;t_{0},s_{0})$
imposes certain constraints on its behavior before the $\theta$.
Namely, if $t_i\ne s_i$ for some $0\le i\le n$, than one can conclude that
an input impulse was obtained just at the end of $t_i$.
In the opposite situation, when $t_i = s_i$, one can conclude that 
in the course of $t_i$ exactly one impulse was obtained from the Poisson stream.
But what do we need in the definition of the 
$P(t_{n+1},s_{n+1}\mid t_{n},s_{n};\ldots;t_{0},s_{0})$,
it is the conditional probability to obtain input
impulses at definite moments after the $\theta$.  
For a Poisson stream this conditional probability does not depend
on conditions before the $\theta$. For example, conditional probability
to obtain the first after $\theta$ impulse at $\theta+t$ 
equals $e^{-\lambda t}\lambda dt$, whatever conditions
are imposed on the stream before the $\theta$.
This proves (\ref{marka}).
\qed


\section{Finding integrals $I_i$ for $P(t_{n+1},\ldots,t_0)$}
\label{sec:Ii}
\begin{figure}
	\includegraphics[height=0.6\textwidth,angle=0]{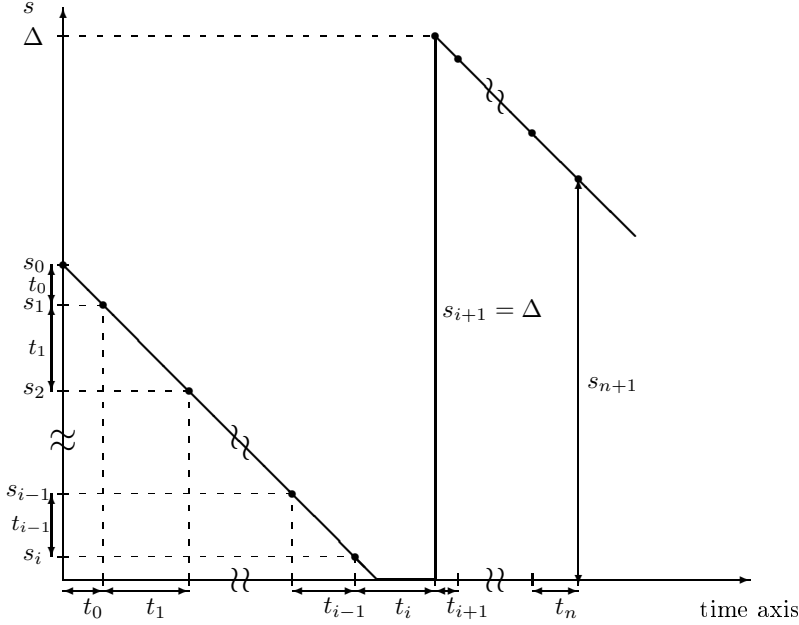}
\caption{
Illustration of relations between $(t_0,\dots,t_n)$ and $(s_0,\dots,s_{n+1})$
contributing to the $I_i$: 
$s_{0}\in\big]\sum_{j=0}^{i-1}t_{j};\sum_{j=0}^{i}t_{j}\big]$, $\sum_{j=0}^n t_j <\Delta$.
The time to live $s_k$ decreases steadily with every output firing
for $k=0,...,i-1$ until it becomes that $s_i<t_i$. 
Then, during the time interval $t_i$ the line discharges
its impulse to BN input, and at the beginning of $t_{i+1}$ starts to convey
the new one with time to live $s_{i+1} = \Delta$.
After that, times to live $s_k$ are again decreased by corresponding $t_k$ with each firing,
$k=i+1,...,n$.
}
\label{fig:ts_scheme}       
\end{figure}

Domain of $s_{0}$ values covered by $I_{i}$, $i=0,\ldots,n$,
corresponds to the scenario, when impulse, which was in the feedback line at the 
beginning of interval $t_{0}$ (with time to live $s_0$), will reach BN during 
interval $t_{i}$, see Figure~\ref{fig:ts_scheme}.
In this process, after each
firing, which starts 
ISI $t_k$, $k\le i$, the time to live of the impulse in the feedback line is decreased
exactly by $t_{k-1}$. This means, that variables of integration $\{s_0,\dots,s_{n+1}\}$, above,
are not actually independent, but must satisfy the following relations:
\begin{equation}
\label{relations1}
	s_k=s_0-\sum\limits_{j=0}^{k-1}t_j,
	\qquad k=1,\dots,i,
\end{equation}
which are also ensured by $\delta$-function in the bottom line of (\ref{P(t,s|t,s)}).
Next to $s_i$ time to live must be equal $\Delta$:
\begin{equation}
\label{relation}
	s_{i+1}=\Delta,
\end{equation}
and this is ensured by $\delta$-function in the top line of (\ref{P(t,s|t,s)}).

The next to $s_{i+1}$ times to live again are decreased by corresponding ISI with
each triggering. Due to (\ref{domain}), this brings about 
another set of relations:
\begin{equation}
\label{relations2}
	s_k=\Delta-\sum\limits_{j=i+1}^{k-1}t_j,
	\qquad k=i+2,\dots,n+1,
\end{equation}
which are again ensured by $\delta$-function in the bottom line of (\ref{P(t,s|t,s)}).
Relations (\ref{relations1}), (\ref{relation}) and (\ref{relations2}) together with limits
of integration over $s_0$ in (\ref{Iidef}) ensure that at $D_1$ the following inequalities hold:
\begin{alignat}{2}
\nonumber
	&s_k>t_k,&\qquad &k=0,\dots,i-1,
\\\nonumber
	&s_i\le t_i,&&
\\
\label{ineqs}
	&s_k>t_k,& \qquad &k=i+1,\dots,n.
\end{alignat}
Inequalities (\ref{ineqs})
allow one to decide
correctly which part of rhs of (\ref{P(t,s|t,s)}) should replace each transition probability
$P(t_{k},s_{k}\mid t_{k-1},s_{k-1})$ in (\ref{Iidef}), and perform all but one 
integration. This gives:
\begin{multline}
	 I_i = \int\limits_{\sum_{j=0}^{i-1}t_j}^{\sum_{j=0}^i t_j} \rmd s_0
	\int\limits_0^{\Delta}\rmd s_1\cdot\ldots\cdot \int\limits_0^{\Delta} \rmd s_{n+1}
\\
         F(t_0\mid s_0) f(s_0)
        \prod_{k=1}^i  F(t_k\mid s_k) \delta(s_k - s_0 + \sum_{j=0}^{k-1}t_j)
\\
 	\times  F(t_{i+1}\mid s_{i+1})\ \delta(s_{i+1} - \Delta)
	\prod_{k=i+2}^{n+1} 
        F(t_k\mid s_k) \delta(s_k-\Delta+\sum_{j=i+1}^{k-1}t_j)
\\
	= \prod\limits_{k=i+1}^{n+1} F(t_{k}\mid \Delta - \sum_{j=i+1}^{k-1} t_j)
	\int\limits_{\sum_{j=0}^{i-1}t_j}^{\sum_{j=0}^i t_j} 
	\prod\limits_{k=0}^{i} F(t_k \mid s_0 - \sum_{j=0}^{k-1}t_j) g(s_0) \rmd s_0,
\\
	\qquad i=0,1,2,\ldots,n.
\end{multline}

The last expression might be obtained as well by means of consecutive substitution
of either top, or bottom line of (\ref{P(t,s|t,s)}) into (\ref{Iidef}), without 
previously discovering (\ref{relations1}) -- (\ref{ineqs}).

Finally, integral $I_{n+1}$ corresponds to the case, when at the beginning of 
interval $t_{n+1}$, the line still keeps the same impulse as at the beginning of $t_{0}$. 
Therefore, $I_{n+1}$ comprises the rest of scenarios contributing to the value of $P(t_{n+1},\ldots,t_{0})$ in (\ref{main}). 
Proceeding as in the preceding terms, the contribution $I_{i+1}$ reads:
\begin{alignat}{1}
\nonumber
	 I_{n+1} &= \int_{\sum_{j=0}^{n}t_{j}}^{\Delta}\rmd s_{0}
	\int_{0}^{\Delta}\rmd s_{1} \ldots \int_{0}^{\Delta} \rmd s_{n+1}
\\\nonumber
 	&\qquad\qquad\qquad\qquad\ \ F(t_{0}\mid s_{0}) f(s_{0})
	\prod_{k=1}^{n+1} F(t_{k}\mid s_{k}) \delta(s_{k}-s_{0}+\sum_{j=0}^{k-1}t_{j}) 
\\\nonumber
	&= \int\limits_{\sum_{j=0}^{n}t_{j}}^{\Delta} 
	\prod\limits_{k=0}^{n+1}	
	F(t_{k}\mid s_{0}-\sum_{j=0}^{k-1}t_{j}) f(s_{0}) \rmd s_{0}
\\
	&= \int\limits_{\sum_{j=0}^{n}t_{j}}^{\Delta} 
	\prod\limits_{k=0}^{n+1}	
	F(t_{k}\mid s_{0}-\sum_{j=0}^{k-1}t_{j}) g(s_{0}) \rmd s_{0}
	+ a\ \prod\limits_{k=0}^{n+1}	
	F(t_{k}\mid \Delta-\sum_{j=0}^{k-1}t_{j}).
\end{alignat}


\section{Continuity of integral factor in (\ref{Ii})}\label{continuity}
 Continuity in $D_1$ of the integral factor
 \begin{equation}
\label{Iicp}
	\int\limits_{\sum_{j=0}^{i-1}t_j}^{\sum_{j=0}^i t_j} 
	\prod\limits_{k=0}^{i} F(t_k \mid s_0 - \sum_{j=0}^{k-1}t_j) g(s_0) \rmd s_0,
	\qquad	i=0,1,\ldots,n,
\end{equation}
  in the expression (\ref{Ii}) can be proven after mathematical simplification. First, notice
  that due to integration domain the following inequalities take place
  $$
 s_0 - \sum_{j=0}^{k-1}t_j>t_k,\quad k=0,1,\dots, i-1,\qquad
 s_0 - \sum_{j=0}^{i-1}t_j<t_i,
 $$
which together with (\ref{P(t|s)sing}) allows to replace (\ref{Iicp}) with
the following
$$
\prod\limits_{k=0}^{i-1}\left(\lambda^2 t_k e^{-\lambda t_k}\right)
\int\limits_{\sum_{j=0}^{i-1}t_j}^{\sum_{j=0}^i t_j} 
\left(1+\lambda s_1\right)
e^{-\lambda s_1} 
P^0\left(t_i-s_1\right)
g(s_0) \rmd s_0,
$$
where $s_1=s_0 - \sum_{j=0}^{i-1}t_j.$
The continuity of the last expression is determined by the continuity of its 
second factor, since the first one is continuous in $\mathbb R^{n+2}$. The second 
factor can be replaced with
\begin{equation}\label{Iicpcp}
\int\limits_0^{t_i} 
(1+\lambda s_0)
e^{-\lambda s_0} 
P^0(t_i-s_0) g\left(s_0 + \sum_{j=0}^{i-1}t_j\right) \rmd s_0.
\end{equation}
after changing the variable of integration. For further simplification of the last 
expression use (\ref{case}), (\ref{domain}) and (\ref{P0}), which gives instead of
(\ref{Iicpcp})

\begin{gather}\nonumber
\int\limits_0^{t_i} 
(1+\lambda s_0)
e^{-\lambda s_0} 
\lambda^2(t_i-s_0) e^{-\lambda(t_i-s_0)}
g\left(s_0 + \sum_{j=0}^{i-1}t_j\right) \rmd s_0=
\\\label{Iicpcpcp1}
=
e^{-\lambda t_i} t_i
\int\limits_0^{t_i}
(1+\lambda s_0)
\lambda^2
g\left(s_0 + \sum_{j=0}^{i-1}t_j\right) \rmd s_0-
\\\label{Iicpcpcp2}
-
e^{-\lambda t_i}
\int\limits_0^{t_i} 
(1+\lambda s_0)
\lambda^2s_0 \ 
g\left(s_0 + \sum_{j=0}^{i-1}t_j\right) \rmd s_0.
\end{gather}
The required continuity of (\ref{Iicp}) is determined by the continuity of 
integral factors in (\ref{Iicpcpcp1}) and (\ref{Iicpcpcp2}). Now, take into account
the explicit expression for $g(s)$, which is found in \cite[Eq. (15)]{BNDiF}, this issue. For our purposes it is enough to know that
$
g(s) = A + B e^{2\lambda s},
$
where $A$ and $B$ are constants. Taking this into account, the integral factor in
(\ref{Iicpcpcp1}) can be replaced with
$$
A\int\limits_0^{t_i}
(1+\lambda s_0)
\lambda^2
 \rmd s_0
+
Be^{2\lambda \sum_{j=0}^{i-1}t_j}
\int\limits_0^{t_i}
(1+\lambda s_0)
\lambda^2
e^{2\lambda s_0} \rmd s_0,
$$
which makes its continuity self-evident. The same is for integral factor in 
(\ref{Iicpcpcp2}).


\end{appendix}

{\small\bf Acknowledgements.}
{\small For numerical simulation, we used utilities from the GNU 
Scien\-ti\-fic Lib\-rary,
which is the free software under GNU General Public Licence, see http://www.gnu.org/software/gsl/.
This work is partially supported by the following projects of the National Academy
of Science of Ukraine: 
(i) Microscopic and phenomenological models of fundamental physical
processes in a micro and macroworld, 
PK N$^{\underbar{\footnotesize o}}$ 0112U000056; 
(ii) Formation of structures in quantum and classical equilibrium
and nonequilibrium systems of interacting particles, 
PK N$^{\underbar{\footnotesize o}}$ 0107U006886.}

\end{document}